\documentclass[a4paper,10pt,twoside]{article}

\newlength{\minipagewidth}
\usepackage{a4wide}
\usepackage{amsmath,amsthm}
\usepackage{amssymb}
\usepackage{graphicx}
\usepackage{color}
\usepackage{epic}
\usepackage{enumerate}

\newcommand{\mC}{\mathrm{C}}
\newcommand\dps {\displaystyle }

\newcommand{\rme}{\mathrm{e}}

\newcommand{\cE}{{\cal E}}

\newcommand{\cP}{{\cal P}}

\newtheorem{theorem}{Theorem}

\newtheorem{remark}{Remark}

\newtheorem{assumption}{Assumption}
\def\sqw{\hbox{\rlap{\leavevmode\raise.3ex\hbox{$\sqcap$}}$%
\sqcup$}}
\def\cqfd{\ifmmode\sqw\else{\ifhmode\unskip\fi\nobreak\hfil
\penalty50\hskip1em\null\nobreak\hfil\sqw
\parfillskip=0pt\finalhyphendemerits=0\endgraf}\fi}
 {\cqfd \vspace{0.2cm}}

\begin{document}


\title{Sampling constraints in average: The example of Hugoniot curves}
\author{Jean-Bernard Maillet$^1$ and Gabriel Stoltz$^{2}$\\
  \footnotesize{1: CEA/DAM, BP 12, 91680 Bruy\'eres-le-Ch\^atel, France.} \\
  \footnotesize{2: Universit\'e Paris Est, CERMICS, Projet MICMAC ENPC -
    INRIA, 6 \& 8 Av. Pascal, 77455 Marne-la-Vall\'ee Cedex 2, France.} \\
}

\maketitle

\begin{abstract}
  We present a method for sampling microscopic configurations
  of a physical system distributed according to a
  canonical (Boltzmann-Gibbs) measure, with a constraint holding in average.
  Assuming that the constraint can be controlled by the volume and/or
  the temperature of the system, and considering
  the control parameter as a dynamical variable, a sampling
  strategy based on a nonlinear stochastic process is
  proposed. Convergence results for this dynamics are proved using
  entropy estimates.
  As an application, we consider the computation of points along the
  Hugoniot curve, which are equilibrium states obtained after
  equilibration of a material heated and compressed by a shock wave.
\end{abstract}


Statistical physics provides a way to obtain macroscopic quantities
starting from systems described at the microscopic level.
In this framework, the state of the system is described by some 
probability measure, the precise choice of the measure depending on the
invariant quantities at hand.
A fundamental statistical
ensemble for instance is the microcanonical ensemble
which is generated by the ergodic limit of the Hamiltonian dynamics, and corresponds to
isolated systems. Another classical ensemble is the canonical
ensemble, for which the temperature is constant. 
In one possible derivation of the associated probability measure, 
the statistical entropy is maximized subject to the constraint that
the average energy of the system is fixed, and the temperature is
related to the Lagrange multiplier of the former constraint~\cite{Balian}.
More generally, it
may be of interest to consider ensembles such that some 
constraint (not only the energy) is satisfied on average. The
challenge is to have an explicit description of the corresponding thermodynamic
ensemble in terms of the associated probability measure.
In an effort to define the
statistical distributions sampled by methods aiming at satisfying
constraints on average, we first reformulate here the problem of
sampling with constraints satisfied in average in the context of the
canonical ensemble.

The paper is organized as follows. In
Section~\ref{sec:mathematical_formulation}, we formulate precisely the
problem of sampling canonical ensembles with a constraint fixed in
average, giving a specific example and recalling some
previous sampling strategies. In Section~\ref{sec:sampling_algo}, we
present a nonlinear stochastic process with a feedback term depending
on the expectation of the constraint, who admits the target measure
as a stationary state. We also prove the convergence to this target
measure using entropy estimates for initial data not too far from the
limiting state (the proofs being presented in
Section~\ref{sec:proofs}). In Section~\ref{sec:numerical_results}, we discuss the
numerical implementation, proposing a single trajectory
discretization, and use the method to compute the Hugoniot curve of Argon.


\section{Mathematical formulation of the problem}
\label{sec:mathematical_formulation}

Consider a microscopic system of $N$ particles in a space of
dimension $d=3$,
described by their positions $q=(q_1,\dots,q_N)$ and momenta
$p=(p_1,\dots,p_N)$, with associated mass matrix $M = {\rm
  Diag}(m_1,\dots,m_N)$, and interacting through a potential $V$.
Periodic boundary conditions are used, and the particles
then stay in a domain $\mathcal{D} = L_x \mathbb{T} \times L_y
\mathbb{T} \times L_z \mathbb{T}$ ($\mathbb{T}$
denoting the one-dimensional torus $\mathbb{R}/\mathbb{Z}$). The volume of the domain
is denoted by $|\mathcal{D}|$. The phase-space $\Omega$ is
\[
\Omega = \mathcal{M} \times \mathbb{R}^{3N} 
= \mathcal{D}^N \times \mathbb{R}^{3N}.
\]
The central quantity describing
the system is the Hamiltonian
\begin{equation}
  \label{eq:hamiltonian}
  H(q,p) = \sum_{i=1}^N \frac{p_i^2}{2m_i} + V(q_1,\dots,q_N).
\end{equation}
The canonical measure associated with the
Hamiltonian~(\ref{eq:hamiltonian}) has a density~\cite{Balian}
\begin{equation}
  \label{eq:canonical_measure}
  \pi_{\mathcal{D},T}(q,p) = \frac{1}{Z_{\mathcal{D},T}}
  \textrm{e}^{-\beta H(q,p)},
  \qquad
  \beta^{-1} = k_{\rm B} T,
\end{equation}
where $k_{\rm B}$ is the Boltzmann
constant, and the partition function $Z_{\mathcal{D},T}$
is a normalization factor so
that~(\ref{eq:canonical_measure}) is indeed a probability measure:
\begin{equation}
  \label{eq:partition_fct}
  Z_{\mathcal{D},T} = \int_\Omega \textrm{e}^{-\beta H(q,p)} \, dq \, dp.
\end{equation}
We have indicated explicitly the dependence of the canonical
measure~(\ref{eq:canonical_measure}) and the partition
function~(\ref{eq:partition_fct}) on the temperature $T$ and
the domain $\mathcal{D}$.
Average thermodynamic properties of the system can be computed as
averages of functions of the microscopic variables
(the so-called observables) with respect to the canonical measure at a temperature $T$
and for a given simulation box $\mathcal{D}$:
\begin{equation}
  \label{eq:canonical_average}
  \langle A \rangle_{\mathcal{D},T} =
  \int_\Omega A(q,p) \, \pi_{\mathcal{D},T}(q,p) \, dq \, dp.
\end{equation}
Canonical simulations are performed at a fixed volume. However, this
is an ill-defined notion since there are infinitely many simulations
domains having the same volume.
In practice, the shape of the simulation
box (for instance cubic) is kept constant, and the
volume is changed using a scaling of the positions $q$. 
We will in any case assume
in the sequel that the knowledge of the volume $|\mathcal{D}|$ is enough
to characterize the domain $\mathcal{D}$.

\subsection{Equilibrium sampling}

In most cases, equilibrium sampling at a fixed temperature and for
a fixed geometry $\mathcal{D}$ is performed
(see~\cite{FrenkelSmit,AT,CLS07} for references on sampling
methods), and the output of the simulation is the average property
$\langle A \rangle_{\mathcal{D},T}$. 
This can be done for instance by means of ergodic results for 
the Langevin dynamics
\begin{equation}
  \label{eq:Langevin}
  \left \{
  \begin{array}{ccl}
    dq_t & = & \dps M^{-1} p_t \, dt, \\ [5pt]
    dp_t & = & \dps -\nabla V(q_t) \, dt - \xi M^{-1} p_t \, dt
    + \sqrt{2 \xi k_{\rm B} T} \, dW_t, \\ [5pt]
  \end{array}
  \right.
\end{equation}
where $\xi > 0$ is physically interpreted as a friction, and 
$W_t$ is a standard $3N$-dimensional Brownian
motion. Under mild assumptions on the potential $V$~\cite{CLS07}, 
\[
\lim_{\tau \to +\infty} \frac1\tau \int_0^\tau A(q_t,p_t) \, dt = 
\langle A \rangle_{\mathcal{D},T} \quad \mathrm{a.s.}
\]
for almost all initial conditions.
When the observable $A$ considered depends only on the positions of
the particles (which is not too restrictive since 
the momentum contribution of many observables can be
explicitly computed), the overdamped 
Langevin dynamics
\[
dq_t  = -\nabla V(q_t) \, dt + \sqrt{2 k_{\rm B} T} \, dW_t,
\]
may be used as well since (again, under some assumptions on the potential~\cite{CLS07})
\[
\lim_{\tau \to +\infty} \frac 1\tau \int_0^\tau A(q_t) \, dt = 
\frac{\dps \int_\mathcal{M} A(q) \, \textrm{e}^{-\beta V(q)} \, dq}
     {\dps \int_\mathcal{M} \textrm{e}^{-\beta V(q)} \, dq} 
     = \langle A \rangle_{\mathcal{D},T} \quad \mathrm{a.s.}
\] 

\subsection{Satisfying constraints in average}

The inverse sampling problem
is sometimes of interest: given some value $A_0$ of an average
property (for instance, some average energy), what should the
temperature and/or the volume of the system be? Answering this
question allows to sample configurations of the system
canonically distributed and satisfying the constraint $A(q,p) =
A_0$ in average (in the sense that $\langle A
\rangle_{\mathcal{D},T} = A_0$).
In the sequel we will focus on constraints depending on $T$,
and therefore drop the mention of the volume in the notation of the
canonical averages when it is not relevant.
Constraints on the volume can be handled similarly. 
Moreover, upon replacing $A$ by $A-A_0$, it can be assumed that $A_0 =
0$, so that the problem under investigation is
\begin{equation}
  \label{eq:problem}
  \fbox{$
    \text{Find } T \text{ such that } \langle A \rangle_{T} = 0.
    $}
\end{equation}
In order to
have a well-defined problem (and possibly upon replacing $A$ by $-A$), 
it is assumed that

\begin{assumption}
  \label{ass:averages}
  There exists an interval $I^A_T = [T^A_\mathrm{min},T^A_\mathrm{max}]$ (with
  $T^A_\mathrm{min} > 0$), a temperature $T^\ast \in
  (T^A_\mathrm{min},T^A_\mathrm{max})$,
  and constants $a, \alpha > 0$ such that 
  \[
  \forall T \in I_T^A, \quad \langle A \rangle_{T} = 0 \Leftrightarrow T = T^\ast,
  \]
  and
  \[
  \forall T \in I_T^A, \quad \alpha \leq 
  \frac{\langle A \rangle_{T} - \langle A \rangle_{T^\ast}}{T - T^\ast} \leq a.
  \]
\end{assumption}

The assumption on the observable is satisfied as
soon as $T \mapsto \langle A \rangle_T$ is locally $\mathrm{C}^1$
around a value $T^*$ such that $\langle A \rangle_{T^\ast} = 0$ 
and $\left. \partial_T \left( \langle A \rangle_T \right
)\right|_{T^\ast} > 0$. An example is illustrated in
Figure~\ref{fig:average_Ac} for the
test case considered in Section~\ref{sec:Hugoniot}.

\subsection{Some examples}

\subsubsection{Hugoniot curves}
\label{sec:Hugoniot_pbm}

The method presented in this paper is illustrated on 
Hugoniot sampling (see Section~\ref{sec:Hugoniot}).
The variations of macroscopic quantities across a shock interface are
governed by the Rankine-Hugoniot relations, which relate the jumps of
the quantities under investigation (pressure, density, velocities) to
the velocity of the shock front.
The third Rankine-Hugoniot conservation law for the Euler
equation governing the hydrodynamic evolution of the fluid reads
(macroscopic quantities are denoted by curly letters)
\begin{equation}
  \label{eq:RH3}
  \cE - \cE_0 - \frac12 (\cP + \cP_0) (\mathcal{V}_0-\mathcal{V}) = 0.
\end{equation}
In this expression, $\cE$ is the internal energy of the fluid, $\cP$ its
pressure, and $\mathcal{V}$ its volume. The subscript $0$ refers to the initial
state, the other quantities are evaluated at a shocked state,
\textit{after equilibration}.
The Hugoniot curve corresponds to all the possible states
satisfying~(\ref{eq:RH3}). In practice, it is computed by considering
shocks of different strengths, inducing various compressions.

A reference temperature $T_0$ and a simulation cell $\mathcal{D}_0$, for instance
$\mathcal{D}_0 = (L_0 \mathbb{T})^3$,
characterize the equilibrium state before the shock. 
In numerical experiments, the compression rate
\[
c = \frac{|\mathcal{D}|}{|\mathcal{D}_0|}
\]
is varied from 1 to some maximal compression rate $0 <
c_\text{max} < 1$, so that $\mathcal{D} = (c^{1/3} L_0 \mathbb{T})^3$
when the compression is isotropic. Since all
macroscopic quantities arising in the hydrodynamic equations are
obtained relying on some local thermodynamic equilibrium
assumption, (\ref{eq:RH3}) can be reformulated at the microscopic
level using statistical mechanics. For a given compression rate~c,
\[
\langle H \rangle_{|\mathcal{D}|,T} - \langle H \rangle_{|\mathcal{D}_0|,T_0} -
  \frac12 (\langle P \rangle_{|\mathcal{D}|,T}
  + \langle P \rangle_{|\mathcal{D}_0|,T_0}) (|\mathcal{D}_0|-|\mathcal{D}|) = 0,
\]
where the pressure observable for a domain $\mathcal{D}$ is 
\[
P(q,p) = \frac{1}{3 |\mathcal{D}|} \sum_{i=1}^N
\frac{p_i^2}{2m_i} - q_i \cdot \nabla_{q_i} V(q).
\]
Notice that we indicate explicitly the dependence on the volumes
for clarity, though the initial and final volumes are given.
The final temperature $T$ is the only unknown, and it satisfies
\[
\left \langle H(q,p) - \langle H \rangle_{|\mathcal{D}_0|,T_0}
  + \frac12 (P(q,p) + \langle P \rangle_{|\mathcal{D}_0|,T_0})
  (1-c) |\mathcal{D}_0| \right \rangle_{|\mathcal{D}|,T} = 0.
\]
Introducing the Hugoniot observable (parametrized by the compression
parameter $c$)
\begin{equation}
  \label{eq:function_hugoniot}
  A_c(q,p) = H(q,p) - \langle H \rangle_{|\mathcal{D}_0|,T_0}
  + \frac12 (P(q,p) + \langle P \rangle_{|\mathcal{D}_0|,T_0})
  (1-c) |\mathcal{D}_0|,
\end{equation}
the Hugoniot problem can be reformulated as
\begin{equation}
  \label{eq:Hugoniostat_problem}
  \text{For a given compression } c_\text{max} \leq c \leq 1,
  \text{ find } T \text{ such that }
  \langle A_c \rangle_{c|\mathcal{D}_0|,T} = 0.
\end{equation}
The compression rate~$c$ parametrizes a curve in the $(P,T)$
diagram, called the Hugoniot curve. Note that this curve does not
correspond to a thermodynamic path.

Since shock waves propagate in one direction (for instance parallel to
the $x$ axis), anisotropic version of the Hugoniot problem are of
interest.
In this case, the compression acts
in the $x$ direction only, and 
$\mathcal{D} = c L_0 \mathbb{T} \times (L_0 \mathbb{T})^2$.
The average pressure $P$ is replaced by the
$P_{xx}$ component of the pressure tensor:
\[
P_{xx}(q,p) = \frac{1}{|\mathcal{D}|} \sum_{i=1}^N
\frac{p_{i,x}^2}{2m_i} - q_{i,x} \nabla_{q_{i,x}} V(q),
\]
where $q_{i,x}, p_{i,x}$ denote respectively the $x$ components of the
position and momenta of the $i$-th particle, and the observable $A_c$ is
replaced by
\[
A_{xx,c}(q,p) = H(q,p) - \langle H \rangle_{|\mathcal{D}_0|,T_0}
  + \frac12 (P_{xx}(q,p) + \langle P_{xx} \rangle_{|\mathcal{D}_0|,T_0})
  (1-c) |\mathcal{D}_0|.
\]
The $xx$ component of the initial pressure tensor
$\langle P_{xx} \rangle_{|\mathcal{D}_0|,T_0}$ may be replaced by
the average pressure $\langle P \rangle_{|\mathcal{D}_0|,T_0}$
when the initial state is isotropic.

\subsubsection{A more general case}

More generally,
Hamiltonians depending on some external parameter$~\lambda$ can be
considered, with
associated observables $A$ controlled by the value of $\lambda$.
For instance, $\lambda$ could be
the intensity of a magnetic field for a spin system, and $A$ the total magnetization.
The problem is then to find $\lambda$ such that
\[
\langle A \rangle_{\mathcal{D},T,\lambda} =
\frac{\dps \int_\Omega A(q,p) \, \textrm{e}^{-\beta H_\lambda(q,p)} \, dq \, dp}
     {\dps \int_\Omega \textrm{e}^{-\beta H_\lambda(q,p)} \, dq \, dp} = 0,
\]
the volume $\mathcal{D}$ and the temperature $T$ being fixed.

\subsection{Previous sampling algorithms}

\subsubsection{Newton's method}
\label{sec:Newton}

An obvious sampling strategy to compute the zero of the function 
$T \mapsto F(T) = \langle A \rangle_T$ 
is to resort to Newton's algorithm. It is
expected that such an approach converges in a few steps if the
function $F$ is well behaved. However,
the derivative $F'(T)$ often
involves the computation of covariances, which are known to require 
long sampling times to be reliably computed. For instance, when trying
to sample configurations with the average energy fixed, $A=H$ and
\[
\partial_T (\langle H \rangle_{T}) = \frac{1}{k_{\rm B} T^2}
(\langle H^2 \rangle_{T}-\langle H \rangle_{T}^2).
\]
A way to bypass this difficulty is to estimate numerically the
derivative by computing averages at two temperatures $T+\Delta T$ and
$T-\Delta T$ around the temperature $T$, and to approximate the
derivative as
\[
\partial_T (\langle A \rangle_{T}) \simeq
\frac{\langle A \rangle_{T+\Delta T}-\langle A \rangle_{T-\Delta T}}{2
    \Delta T}.
\]
This method is robust, and is indeed used in Monte-Carlo studies of
the Hugoniot problem~\cite{BR03}. It requires however very carefully
converged canonical samplings in order to have a numerical
estimate of the derivative not polluted by sampling errors.

\subsubsection{New thermodynamic ensembles}

It would be better to have a more automatic procedure,
consisting for instance of a single molecular dynamics
trajectory. In this case, convergence needs only be checked once, and
the simulation can be runned longer if needed.
This was the motivation for the Hugoniostat
method~\cite{MMSRLGH00}, which can be extended to any constraint whose
values are controlled by the temperature $T$.
The idea of the method is to satisfy
approximately the constraint $A(q,p) = 0$ at all times.
To this end, a convenient dynamics in the Nos\'e-Hoover 
fashion~\cite{Nose84,Hoover85} is
postulated:
\begin{equation}
  \label{eq:NH_feedback}
  \left\{
  \begin{array}{ccl}
    \dot{q} & = & M^{-1} p,  \\
    \dot{p} & = & -\nabla V(q) - \xi p, \\ [5pt]
    \dot{\xi} & = & \dps \nu^2 \frac{A(q,p)}{A_{\rm ref}}, \\
  \end{array}
  \right.
\end{equation}
where $\nu > 0$ is homogeneous to a frequency, and $A_{\rm ref}$ is
some reference value of the observable.
When the instantaneous value of the observable is lower than the
target value $0$, $\xi$ decreases and the friction is reduced, so
that the (kinetic) temperature in the system, and therefore the
average value of the observable, can increase; while the
friction increases for too large values of the observable, and so,
energy is removed from the system and the temperature decreases.
This feedback process ensures indeed that values of the observable around the
target value are sampled, though it is unclear how the temperature of
the system is defined. For
computational purposes, the temperature is defined as the average
kinetic temperature along the trajectory.

The numerical results obtained with the
dynamics~\eqref{eq:NH_feedback} are correct for the Hugoniot
problem since they are in agreement with numerical results from
shock wave simulations, see~\cite{MMSRLGH00} for more precisions.
However, the question of the choice of the frequency $\nu$
in~(\ref{eq:NH_feedback}) remains, as usual for Nos\'e-like
dynamics. More importantly, from a fundamental viewpoint, an
unpleasant feature is that the thermodynamic ensemble ({\it i.e.}
the probability measure on phase space) associated with the
dynamics used is not known. It may be defined as the ergodic limit
of the dynamics, for a given set of parameters and initial
conditions. Even if this is possible, it is unclear whether
the limiting measure depends on the parameters used and on the
initial conditions, or not. The problem remains even if the
Nos\'e-Hoover part of the above dynamics is replaced by a Langevin
dynamics (for which ergodicity results are usually easier to
state) or any dynamics ergodic for the canonical measure
(see~\cite{CLS07} for references on the convergence properties of
some sampling methods). For instance, the usual Langevin dynamics
\eqref{eq:Langevin} at temperature $T$
may be extended to the case of constraints satisfied in average by
considering the temperature as a dynamical variable:
\begin{equation}
  \label{eq:Langevin_feedback}
  \left\{
  \begin{array}{ccl}
    dq_t & = & M^{-1} p_t \, dt,  \\
    dp_t & = & -\nabla V(q_t) \, dt - \xi M^{-1} p_t \, dt + \sqrt{2
    \xi k_{\rm B} T_t} \, dW_t, \\ [5pt]
    dT_t & = & -\dps \nu \frac{A(q_t,p_t)}{A_{\rm ref}} \,
    T_\text{ref} \, dt, \\
  \end{array}
  \right.
\end{equation}
where $\nu > 0$ is again some frequency, and $T_\text{ref}$,
$A_{\rm ref}$ are reference values of the temperature and the observable respectively.
When the instantaneous value of the observable is too low, the
temperature is increased, while it is decreased when it is too
large. For computational purposes, the temperature may be
defined as the time average of $T_t$ over a trajectory, though, as for
the dynamics~(\ref{eq:NH_feedback}), the meaning of this quantity is
unclear since there is no obvious definition
of the thermodynamic temperature within the ensemble
generated by the dynamics~(\ref{eq:Langevin_feedback}). Therefore,
\eqref{eq:Langevin_feedback} is not satisfactory.


\section{A nonequilibrium sampling strategy}
\label{sec:sampling_algo}

\subsection{General algorithm}
\label{sec:derivation_dynamics}

To solve the sampling problem~\eqref{eq:problem}, 
we consider the temperature as a variable, so that the
system is described by the variables $(q,p,T)$.
The idea is that the update of the current configuration $(q,p)$ is governed
by a dynamics consistent with canonical sampling at the instantaneous temperature
$T(t)$ (such as Nos\'e-Hoover methods~\cite{Nose84,Hoover85}, 
Metropolis-Hastings algorithms~\cite{MRRTT53,hastings70}, Langevin or overdamped
Langevin dynamics,...), while the temperature is updated depending on
the current estimate of $\langle A \rangle_{T(t)}$.
For instance, consider the following system when the underlying dynamics is of
overdamped Langevin type:
\begin{equation}
  \label{eq:adaptive_dynamics_SDE}
  \left \{
  \begin{array}{ccl}
    dq_t  & = & \dps -\nabla V(q_t) \, dt + \sqrt{2 k_{\rm B} T(t)} \, dW_t, \\ [5pt]
    T'(t) & = & \dps - \gamma \, \mathbb{E}(A(q_t)), \\
  \end{array}
  \right.
\end{equation}
or
\begin{equation}
  \label{eq:adaptive_dynamics_SDE_Langevin}
  \left \{
  \begin{array}{ccl}
    dq_t & = & \dps M^{-1} p_t \, dt, \\ [5pt]
    dp_t & = & \dps -\nabla V(q_t) \, dt - \xi M^{-1} p_t \, dt
    + \sqrt{2 \xi k_{\rm B} T(t)} \, dW_t, \\ [5pt]
    T'(t) & = & \dps - \gamma \, \mathbb{E}(A(q_t,p_t)), \\
  \end{array}
  \right.
\end{equation}
when the underlying dynamics is of Langevin type.
In both cases, $\gamma > 0$ determines the time scale of the
temperature feedback term, and $W_t$ is a standard $3N$-dimensional Brownian motion.
To prove mathematical results, it is convenient to use
(elliptic) overdamped Langevin dynamics,
since the associated Fokker-Planck
equation describing the evolution of the law of the process $q_t$ is
parabolic. However, we will use the 
hypoelliptic Langevin dynamics in the numerical
applications presented in Section~\ref{sec:numerical_results} since
this dynamics is usually a more efficient sampling device~\cite{CLS07}.
When the overdamped Langevin dynamics is used, 
the canonical measure $\mu_{T}$ in position space has the density
\[
\mu_{T}(q) = \frac{1}{Z_T}
\exp\left( -\frac{V(q)}{k_\mathrm{B} T}\right),
\qquad 
Z_{T} = \int_\mathcal{M} \exp\left( 
-\frac{V(q)}{k_\mathrm{B} T}\right) \, dq,
\]
and averages with respect to the canonical measure $\mu_T$ are still denoted
$\dps \langle A \rangle_T = \int_\mathcal{M} A(q) \, \mu_{T}(q) \, dq$.

\subsubsection{Motivation}

The dynamics \eqref{eq:adaptive_dynamics_SDE} 
and~\eqref{eq:adaptive_dynamics_SDE_Langevin} can be motivated as
follows. If the configurations followed adiabatically the temperature
changes, the positions $q_t$ (and possibly the momenta $p_t$) 
would be distributed canonically at
the temperature $T(t)$ at all times, and 
\[
T'(t) = - \gamma \, \langle A \rangle_{T(t)}. 
\]
Assumption~\eqref{ass:averages} then shows that
$T(t) \to T^\ast$. Indeed, starting from a temperature
$T(0)$ in the interval $I_T^A$
of Assumption~\ref{ass:averages}, 
it is easily seen that $t \mapsto (T(t)-T^\ast)^2$
is a decreasing function: if $T(t) \in I_T^A$, then
\begin{eqnarray*}
\partial_t \left [ \frac12 (T(t)-T^\ast)^2 \right ] 
& = & -\gamma \langle A \rangle_{T(t)} (T(t)-T^\ast)
= -\gamma (\langle A \rangle_{T(t)} - \langle A \rangle_{T^\ast})
(T(t)-T^\ast) \\
& \leq & -\gamma \alpha (T(t)-T^\ast)^2 \leq 0,
\end{eqnarray*}
and so, $T(t) \in I_T^A$ whenever $T(0) \in I_T^A$.
Suppose for instance $T(0) > T^\ast$. Then $T(t) \geq T^\ast$ at all
times and, with Assumption~\ref{ass:averages},
\[
-\gamma a (T(t)-T^\ast) \leq T'(t) = -\gamma (\langle A \rangle_{T(t)}
- \langle A \rangle_{T^\ast}) \leq -\gamma \alpha (T(t)-T^\ast).
\]
Gronwall's lemma then implies
\[
T^\ast + (T(0)-T^\ast) \, \rme^{-a \gamma t} \leq 
T(t) \leq T^\ast + (T(0)-T^\ast) \, \rme^{-\alpha \gamma t},
\]
which shows the convergence $T(t) \to T^\ast$ as $t \to +\infty$.

Of course, the positions are not canonically distributed at all times.
The hope is that, even if equilibrium is not
maintained at all times, the dynamics can still converge if the
typical time arising in the temperature update is small enough
compared to the typical relaxation time of the dynamics, so that the
spatial relaxation happens gradually, as $T(t)$ approaches $T^\ast$. This
statement is made precise in Section~\ref{sec:cv_results}
(see Theorems~\ref{thm:LSIcv} and~\ref{thm:Poincare_cv}).
From a practical viewpoint, the interest of such a strategy is that,
if the dynamics is started off equilibrium (for instance at a
temperature much larger than the target temperature $T^\ast$), then it
is not necessary to fully relax the system at the starting temperature (as is the
case for instance for the Newton method presented in Section~\ref{sec:Newton}), and
so, computational savings may be anticipated.
The convergence proof is however only valid for initial data not too
far from the stationary state. This is related to the fact that the
temperature has to remain positive, and so, it is not
possible in general to start arbitrarily far from equilibrium. 

\begin{remark}[Possible extensions]
\label{rmk:extensions}
A  nonlinear temperature feedback may be considered as well: 
\begin{equation}
  \label{eq:temperatureNLfeedback}
  T'(t) = - \gamma f\left( \mathbb{E}(A(q_t)) \right),
\end{equation}
under some assumptions on the function $f$ 
-- see Remark~\ref{remark:NLfeedback} for more precisions.

It is also possible to use a time varying parameter $\gamma(t)$
for the temperature update,
for instance smaller at the beginning of the simulation, and then
larger once the equilibrium state is approached. The proofs presented
below may be extended to this case 
provided $\gamma(t)$ remains in a properly chosen interval around an
admissible value of $\gamma$ (as given by Theorems~\ref{thm:LSIcv}
and~\ref{thm:Poincare_cv}).
\end{remark}

\subsubsection{Partial differential equation reformulation}

The nonlinear partial differential equation (PDE) 
reformulation of the dynamics~\eqref{eq:adaptive_dynamics_SDE} is
\begin{equation}
  \label{eq:adaptive_dynamics_PDE}
  \left \{
  \begin{array}{ccl}
    \partial_t \psi  & = & \dps k_{\rm B} T(t) \, \nabla \cdot \left [ \mu_{T(t)}
    \nabla \left ( \frac{\psi}{\mu_{T(t)}}\right ) \right ]
    = k_{\rm B} T(t) \, \Delta \psi + \nabla \cdot ( \psi \nabla V), \\ [10pt]
    T'(t) & = & \dps - \gamma \int_{\mathcal{M}} A(q) \psi(t,q) \, dq, \\
  \end{array}
  \right.
\end{equation}
where the periodic function $q \in \mathcal{M} \mapsto \psi(t,q)$ 
is the law of the process $q_t$ at time $t$. 

The PDE \eqref{eq:adaptive_dynamics_PDE} clearly admits
$(T^\ast,\mu_{T^\ast})$ as a stationary solution. 
Since the first equation of \eqref{eq:adaptive_dynamics_PDE} is of parabolic type, 
standard techniques may be used to prove the (short time) 
existence of a unique solution, for initial 
conditions smooth enough, and under the following smoothness
assumptions on the potential and the observable:

\begin{assumption}
  \label{ass:global_assumption}
  The observable $A \in \mathrm{C}^3(\mathcal{M})$ and $V \in \mathrm{C}^2(\mathcal{M})$.
\end{assumption}  

In particular, the observable is bounded, which is important 
to ensure that the temperature remains positive. Indeed,
if this was not the case, it would be possible to obtain negative
temperatures arbitrarily fast
by concentrating the initial density $\psi^0$ around
singularities of the observable. The boundedness of $A$ ensures some
delay in the feedback.

\begin{theorem}
  \label{thm:ExistenceUniqueness}
  When Assumption~\ref{ass:global_assumption} holds, and
  for a given initial condition $(T^0,\psi^0)$, with $T^0 > 0$ and
  \[
  \psi^0 \in \mathrm{H}^2(\mathcal{M}), 
  \qquad 
  \psi^0 \geq 0, 
  \qquad
  \int_\mathcal{M} \psi^0 = 1,
  \]
  there exists a time $\dps \tau \geq \frac{T^0}{2\gamma \| A \|_\infty}
  > 0$ such that \eqref{eq:adaptive_dynamics_PDE} has a unique solution
  $(T,\psi)$ with $T \in \mathrm{C}^1([0,\tau],\mathbb{R})$ and 
  $\psi \in \mathrm{C}^0([0,\tau],\mathrm{H}^2(\mathcal{M}))$.
  Moreover, $\psi \geq 0$, and $\psi > 0$ when $\psi^0 > 0$.
\end{theorem}

The proof is presented in Section~\ref{sec:proof_existence}.
The regularity assumptions on $\psi^0$ is not too
restrictive since an arbitrary initial condition $\widetilde{\psi}^0 \in
\mathrm{L}^2(\mathcal{M})$ can be evolved using an overdamped Langevin
dynamics at $T^0$ for some time $\tau_{\rm init} > 0$ before turning
on the temperature feedback term. Then, the resulting distribution $\psi^0 =
\widetilde{\psi}(\tau_{\rm init},\cdot)$ is smooth thanks to the
regularizing properties of the parabolic Fokker-Planck equation.
To obtain the uniqueness and existence of the solution at all times, it
is necessary to show that the temperature remains isolated from 0.
This is proved in Section~\ref{sec:cv_results} together with the
convergence to the stationary state, under certain conditions on the initial entropy 
(see Theorems~\ref{thm:LSIcv} and~\ref{thm:Poincare_cv}).

\subsection{Convergence results for initial conditions close to the
  fixed-point}
\label{sec:cv_results}

To study the convergence of the nonlinear 
dynamics~\eqref{eq:adaptive_dynamics_PDE} under its PDE form, entropy
methods can be used (see for instance the review papers
\cite{GZ03,AMTU01}). 
The total entropy of the system is the sum of a spatial entropy
$E(t)$, related to the distribution of configurations at time $t$, and
a temperature term:
\[
\mathcal{E}(t) = E(t) + \frac12 (T(t)-T^\ast)^2,
\]
with
\[
E(t) = \int_{\mathcal{M}} h\left(f\right) \mu_{T(t)}, 
\qquad 
f = \frac{\psi}{\mu_{T(t)}}.
\]
Notice that the spatial entropy $E(t)$ measures the distance of the
law of the process $\psi$ to the instantaneous equilibrium measure
$\mu_{T(t)}$, and not to some fixed reference measure. 
Two classical
choices for the spatial entropy 
are the relative entropy distance (which corresponds to 
$h(x) = x \ln x - x +1 \geq 0$) and $\mathrm{L}^2$ distances ($\dps h(x)=\frac12(x-1)^2$).
We present convergence results for both cases (see
Theorems~\ref{thm:LSIcv} and~\ref{thm:Poincare_cv}), which are very
similar in spirit, but involve different functional settings for the
initial conditions. In particular, the conditions on the initial
condition are more stringent in Theorem~\ref{thm:Poincare_cv}, but the
assumptions to be verified by the potential are less demanding.

The convergence is ensured provided $\mathcal{E}(t) \to 0$ as $t \to
+\infty$. This implies in particular $T(t) \to T^\ast$, and also
\[
\| \psi(t,\cdot) - \mu_{T^\ast} \|_{\mathrm{L}^1}
\leq 
\| \psi(t,\cdot) - \mu_{T(t)} \|_{\mathrm{L}^1} + 
\| \mu_{T(t)}  - \mu_{T^\ast} \|_{\mathrm{L}^1}
\leq c_h \, E(t) + \| \mu_{T(t)}  - \mu_{T^\ast}
\|_{\mathrm{L}^1} \to 0,
\]
where the constant $c_h$ depends on the chosen entropy density $h$
(using a Csiz\'ar-Kullback inequality for relative entropy estimates,
and a Cauchy-Schwarz inequality for $\mathrm{L}^2$ entropies).
The strategy of the proof is to obtain a Gronwall inequality for
$\mathcal{E}(t)$. This ensures also that the temperature is well
defined (in particular, $T(t) > 0$ at all times). Indeed, 
if the total entropy is decreasing, then $|T(t)-T^\ast|$ is
bounded by $\sqrt{2\mathcal{E}(0)}$, and the temperature remains
isolated from 0 for initial data whose total entropy is small enough.

In the proofs presented in Section~\ref{sec:proofs}, estimates on the 
derivative of the spatial entropy are a key element. Since
\[
\partial_t \mu_{T(t)} = \frac{T'(t)}{k_\mathrm{B} T(t)^2} \left(
V - \left \langle V \right \rangle_{T(t)} \right ) \, \mu_{T(t)},
\]
and thus
\[
\partial_t f = \frac{\partial_t \psi}{\mu_{T(t)}} - \frac{T'(t)}{k_\mathrm{B} T(t)^2} \left(
V - \left \langle V \right \rangle_{T(t)} \right ) \, f,
\]
it holds
\begin{eqnarray*}
\frac{dE(t)}{dt} & = & \int_{\mathcal{M}} h'(f) \, \partial_t f
\, \mu_{T(t)} + \int_{\mathcal{M}} h(f) \partial_t \mu_{T(t)} \\ 
& = & \int_{\mathcal{M}} h'(f) \, \partial_t \psi
+ \frac{T'(t)}{k_\mathrm{B} T(t)^2}
\int_{\mathcal{M}} \left [ h(f) - h'(f) \, f \right ]\, \left(
V - \left \langle V \right \rangle_{T(t)} \right ) \, \mu_{T(t)}\\
& = & k_{\rm B} T(t) \, 
\int_{\mathcal{M}} h'(f) \, \nabla \cdot \left [ \mu_{T(t)}
  \nabla \left ( \frac{\psi}{\mu_{T(t)}}\right ) \right ] 
+ \frac{T'(t)}{k_\mathrm{B} T(t)^2}
\int_{\mathcal{M}} \left [ h(f) - h'(f) \, f \right ]\, \left(
V - \left \langle V \right \rangle_{T(t)} \right ) \, \mu_{T(t)}\\
& = & - k_{\rm B} T(t) \, \int_{\mathcal{M}} h''(f) \, \left | 
\nabla f \right |^2 \, \mu_{T(t)}
+ \frac{T'(t)}{k_\mathrm{B} T(t)^2}
\int_{\mathcal{M}} \left [ h(f) - h'(f) \, f \right ]\, \left(
V - \left \langle V \right \rangle_{T(t)} \right ) \, \mu_{T(t)}.\\
\end{eqnarray*}
In view of Theorem~\ref{thm:ExistenceUniqueness}, the above
manipulations (in particular the integrations by parts) can be
performed, at least for $t$ small enough.
The proofs of convergence are based on the following 
strategy: The first term in the above equality is bounded by $-\rho
E(t)$ for some $\rho > 0$ using a functional inequality (logarithmic
Sobolev inequality (LSI) or a Poincar\'e inequality, depending on the
functional setting), while the remainder is shown to be small for
$\gamma$ small enough thanks to the temperature derivative factor.

\subsubsection{Relative entropy estimates}

This case corresponds to the choice $h(x) = x \ln x - x +1$.
In addition of Assumption~\ref{ass:global_assumption}, we suppose that 

\begin{assumption}
  \label{ass:LSIcase}
  There exists an interval $I^\mathrm{LSI}_T = 
  [T^\mathrm{LSI}_\mathrm{min},T^\mathrm{LSI}_\mathrm{max}]$ (with
  $T^\mathrm{LSI}_\mathrm{min} > 0$ and $T^\mathrm{LSI}_\mathrm{min} <
  T^\ast < T^\mathrm{LSI}_\mathrm{max}$) such that the family of measures $\{ \mu_T \}_{T
    \in I^\mathrm{LSI}_T}$ satisfies a logarithmic Sobolev inequality (LSI) with a uniform
  constant $1/\rho$, namely
  \[
  \int_\mathcal{M} h(f) \mu_{T} \leq \frac{1}{\rho} 
  \int_\mathcal{M} \frac{|\nabla f|^2}{f} \, \mu_T.
  \]
\end{assumption}

A LSI holds for instance in the following cases:
when the potential~$V$ satisfies a strict convexity 
condition of the form ${\rm Hess}(V) \geq \nu \,
{\rm Id}$ with $\nu > 0$
(as a special case of the Bakry and Emery criterion~\cite{BE85}), 
or when the measure $\mu_T$ is a tensorization of measures
satisfying a LSI (see Gross~\cite{Gross75}).
Moreover, when a LSI with constant~$\rho$ is satisfied by $Z_V^{-1} \, {\rm
  e}^{-V(q)} \, dq$, then $Z_{V+W}^{-1} \, {\rm e}^{-(V(q)+W(q))} \, dq$
(with $W$ bounded) satisfies a LSI with constant $\tilde{\rho} = \rho \, {\rm e}^{\inf W -
  \sup W}$.
This property expresses some 
stability with respect to bounded pertubations (see Holley and Stroock~\cite{HS87}).
In particular, a LSI is verified for the canonical measure associated
with smooth potentials on a compact state space (as is the case
here). The uniformity of the
constant can be ensured by restricting the temperatures to a finite
interval isolated from 0. 

\begin{theorem}
  \label{thm:LSIcv}
  Consider a potential $V$ and an observable $A$ such that
  Assumptions~\ref{ass:averages} and~\ref{ass:LSIcase} hold, 
  and an initial data $(T^0,\psi^0)$ with $\psi^0 \in
  \mathrm{H}^2(\mathcal{M})$, $\psi^0 \geq 0$, 
  $\dps \int_\mathcal{M} \psi^0 = 1$, and associated entropy 
  $\mathcal{E}(0) \leq \mathcal{E}^\ast$, where 
  \[
  \mathcal{E}^\ast = \inf \left \{ \frac12 (T_\mathrm{min}^A-T^\ast)^2,
  \frac12 (T_\mathrm{max}^A-T^\ast)^2, 
  \frac12 (T_\mathrm{min}^\mathrm{LSI}-T^\ast)^2,
  \frac12 (T_\mathrm{max}^\mathrm{LSI}-T^\ast)^2
  \right \}.
  \]
  Then, there exists $\gamma_0 > 0$ such that, for all $0 < \gamma \leq
  \gamma_0$, \eqref{eq:adaptive_dynamics_PDE} has a unique solution 
  $(T,\psi) \in \mathrm{C}^1([0,\tau],\mathbb{R}) \times 
  \mathrm{C}^0([0,\tau],\mathrm{H}^2(\mathcal{M}))$ for all $\tau \geq 0$,
  and the entropy converges exponentially fast to zero: There
  exists $\kappa > 0$ (depending on $\gamma$) such that 
  \[
  \mathcal{E}(t) \leq \mathcal{E}(0) \exp(-\kappa t).
  \] 
  In particular, the temperature remains positive at all times: $T(t)
  \geq \min(T_\mathrm{min}^\mathrm{LSI},T_\mathrm{min}^A) > 0$, and it
  converges exponentially fast to $T^*$.
\end{theorem}

\begin{remark}
The proof shows that the convergence rate $\kappa$ is larger 
when (i) $\mathcal{E}(0)$ is lower
(the dynamics starts closer from the fixed point and/or closer from a
spatial local equilibrium); (ii) $\alpha$ is larger (the slope of
the function $T \mapsto \langle A \rangle_T$ is
steeper around the target value $T^\ast$); (iii) $\rho$ is larger (the
relaxation of the spatial distribution of configurations at a fixed
temperature happens faster).
\end{remark}

\subsubsection{$\mathrm{L}^2$ estimates}

This case corresponds to the choice $\dps h(x) = \frac12(x-1)^2$. 
In this section, in addition of Assumption~\ref{ass:global_assumption} (and instead of
Assumption~\ref{ass:LSIcase}), we suppose that 

\begin{assumption}
  \label{ass:Poincare_case}
  There exists an interval $I^\mathrm{Poincare}_T = 
  [T^\mathrm{Poincare}_\mathrm{min},T^\mathrm{Poincare}_\mathrm{max}]$ (with
  $T^\mathrm{Poincare}_\mathrm{min} > 0$ and $T^\mathrm{Poincare}_\mathrm{min} <
  T^\ast < T^\mathrm{Poincare}_\mathrm{max}$) such that 
  the family of measures $\{ \mu_T \}_{T
    \in I^\mathrm{Poincare}_T}$ satisfies a Poincar\'e inequality with a uniform
  constant $1/\rho$, namely
  \[
  \int_\mathcal{M} h(f) \mu_{T} \leq \frac{1}{\rho} 
  \int_\mathcal{M} |\nabla f|^2 \, \mu_T.
  \]
\end{assumption}

Recall that LSI imply Poincar\'e inequalities (see for instance \cite{AMTU01}), so that the second part
of Assumption~\ref{ass:Poincare_case} is less demanding than
the second part of Assumption~\ref{ass:LSIcase}, and is verified for instance for 
smooth potentials on compact state spaces.
A result analogous to Theorem~\ref{thm:LSIcv} can be stated upon
defining another bound on the initial entropy (recall that bounds
in $\mathrm{L}^2$ entropy are more demanding than bounds in relative entropy~\cite{AMTU01}).

\begin{theorem}
  \label{thm:Poincare_cv}
  Consider a potential $V$ and an observable $A$ such that
  Assumptions~\ref{ass:averages} and~\ref{ass:Poincare_case} hold, 
  and an initial data $(T^0,\psi^0)$ with $\psi^0 \in
  \mathrm{H}^2(\mathcal{M})$, $\psi^0 \geq 0$, 
  $\dps \int_\mathcal{M} \psi^0 = 1$, and associated entropy 
  $\mathcal{E}(0) \leq \mathcal{E}^\ast$, where 
  \[
  \mathcal{E}^\ast = \inf \left \{ \frac12 (T_\mathrm{min}^A-T^\ast)^2,
  \frac12 (T_\mathrm{max}^A-T^\ast)^2, 
  \frac12 (T_\mathrm{min}^\mathrm{Poincare}-T^\ast)^2,
  \frac12 (T_\mathrm{max}^\mathrm{Poincare}-T^\ast)^2
  \right \}.
  \]
  Then, there exists $\gamma_0 > 0$ such that, for all $0 < \gamma \leq
  \gamma_0$, 
  \eqref{eq:adaptive_dynamics_PDE} has a unique solution 
  $(T,\psi) \in \mathrm{C}^1([0,\tau],\mathbb{R}) \times 
  \mathrm{C}^0([0,\tau],\mathrm{H}^2(\mathcal{M}))$ for all $\tau \geq 0$,
  and the entropy converges exponentially fast to zero: There
  exists $\kappa > 0$ (depending on $\gamma$) such that 
  \[
  \mathcal{E}(t) \leq \mathcal{E}(0) \exp(-\kappa t).
  \] 
  In particular, the temperature remains positive at all times: $T(t)
  \geq \min(T_\mathrm{min}^\mathrm{Poincare},T_\mathrm{min}^A) > 0$,
  and it converges exponentiallu fast to $T^*$.
\end{theorem}


\section{Numerical results}
\label{sec:numerical_results}

\subsection{Single trajectory discretization of the dynamics}
\label{sec:single_traj_discretization}

\begin{figure}[h]
\center
\includegraphics[width=7.2cm]{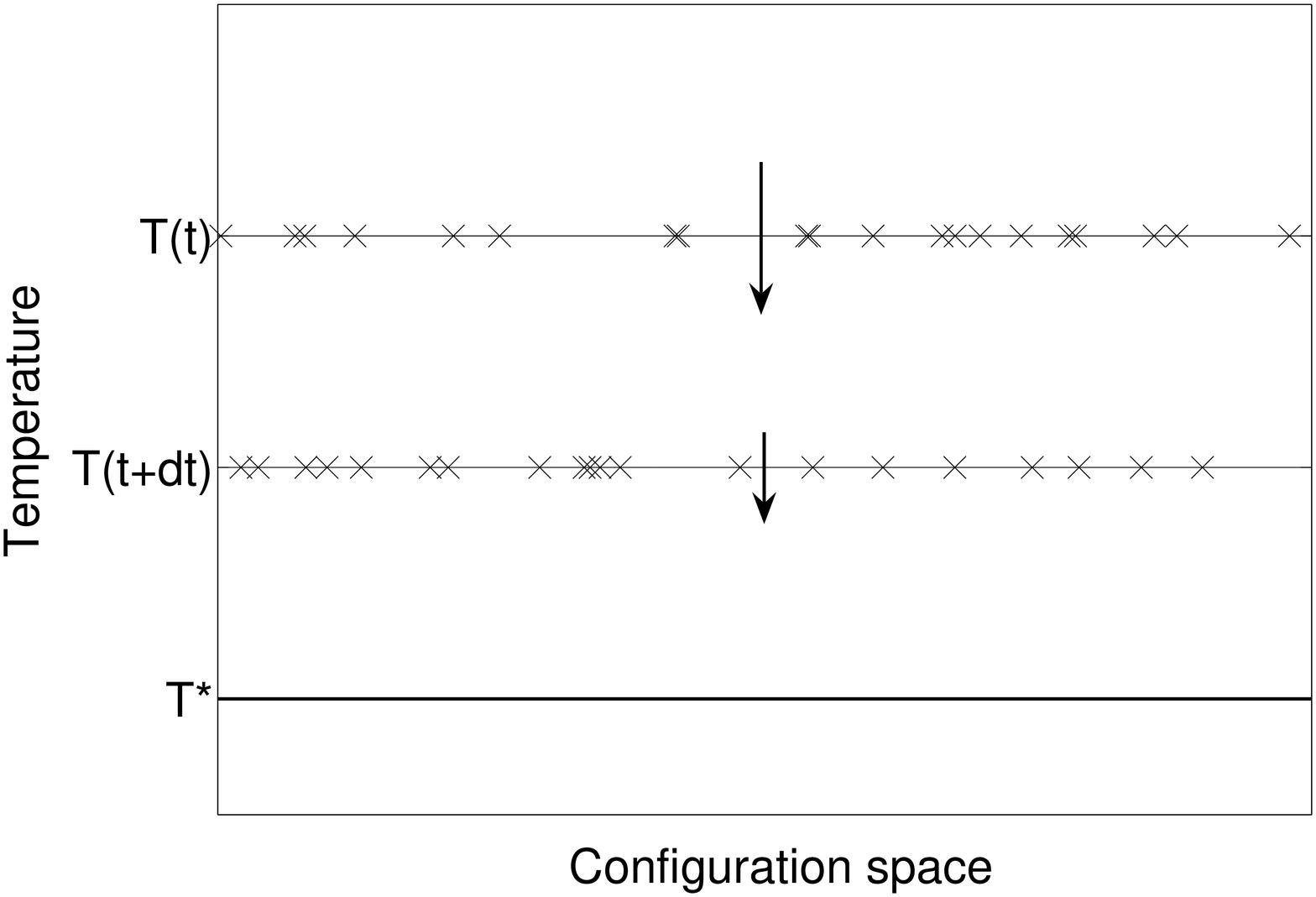}\hfill
\includegraphics[width=7.2cm]{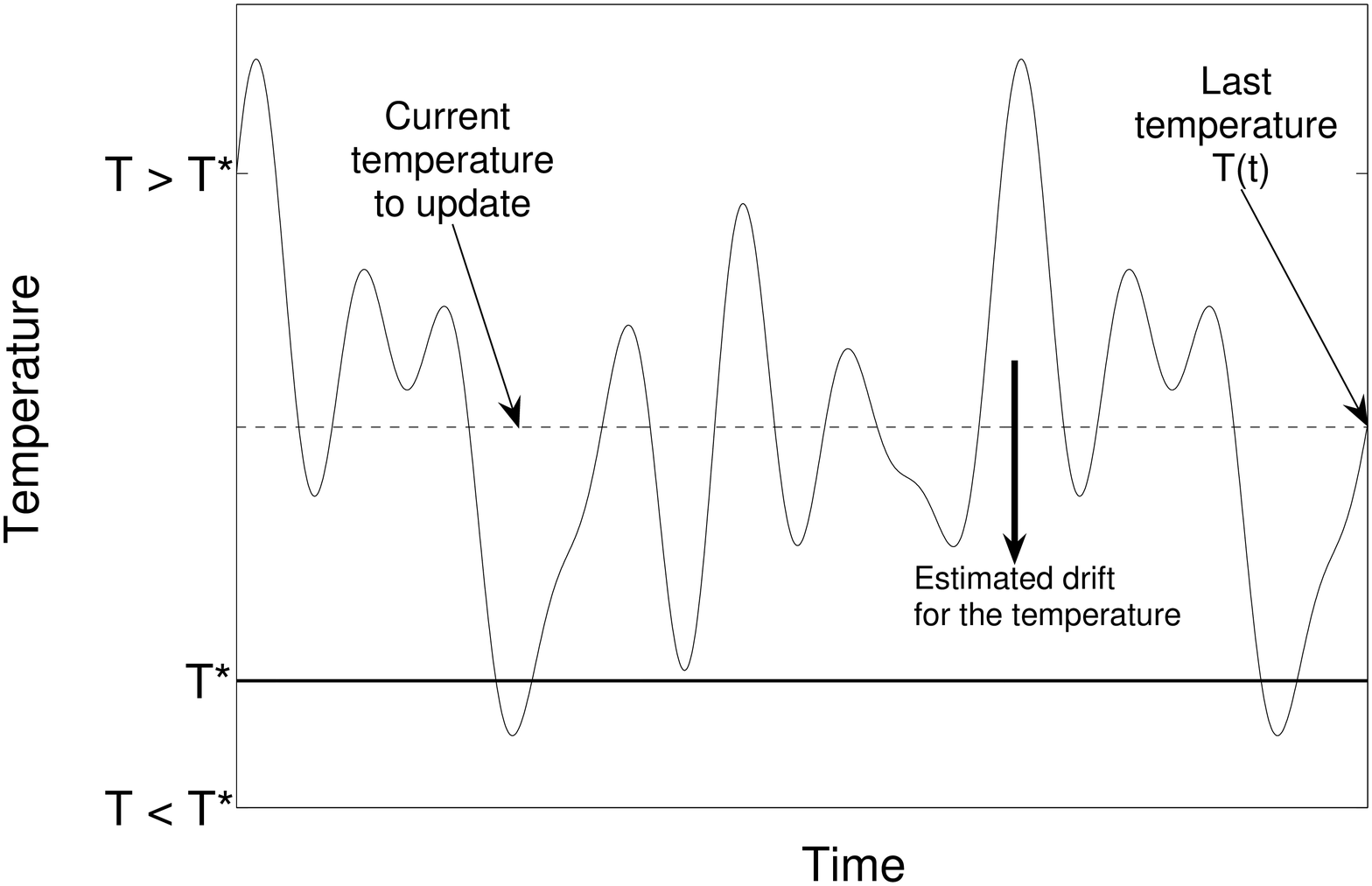}
\caption{\label{fig:single_traj_discretization}
  Left: Discretization using several replicas of the system
  interacting only through the temperature update term. All the
  replicas are at the same temperature, and estimates of the
  temperature update term are obtained from the instantaneous
  distribution of configurations. Right: Single replica
  discretization. The estimate of the update term at a given
  temperature is done in analogy with the many replica case, 
  using the previous configurations sampled at the
  same temperature.
}
\end{figure}

We present in this section a possible discretization of the dynamics 
\eqref{eq:adaptive_dynamics_SDE_Langevin}. Recall however that the
general discretization method presented here may be adapted to any
dynamics consistent with the canonical ensemble.
A first obvious strategy is to approximate the expectation by some
empirical average over $K$ replicas of the system:
\[
\forall \ 1 \leq k \leq K, \quad \left \{ \begin{array}{cl}
  dq^k_t = & \dps M^{-1} p^k_t \, dt, \\ [5pt]
  dp^k_t = & \dps -\nabla V(q^k_t) \, dt - \xi M^{-1} p^k_t \, dt
  + \sqrt{2 \xi k_{\rm B} T(t)} \, dW^k_t, \\ 
\end{array} \right. 
\]
(where $(W_t^k)_{k=1,\dots,K}$ are standard independent
$3N$-dimensional Brownian motions) 
interacting only through the update of their common temperature:
\[
T'(t) = -\dps \frac{\gamma}{K} \sum_{k=1}^K A(q_t^k,p_t^k).
\]
However, it is often more convenient from a practical
perspective to replace
the average over many replicas simulated in parallel by an average
over a single realization. To this end, the
dynamics~(\ref{eq:adaptive_dynamics_SDE_Langevin}) is approximated by the following
dynamics:
\begin{equation}
  \label{eq:single_replica_dynamics}
  \left \{
  \begin{array}{cl}
    dq_t = & \dps M^{-1} p_t \, dt, \\ [5pt]
    dp_t = & \dps -\nabla V(q_t) \, dt - \xi M^{-1} p_t \, dt
    + \sqrt{2 \xi k_{\rm B} T_t} \, dW_t, \\ [5pt]
    dT_t = & \dps - \gamma \left ( \frac{\dps \int_0^t A(q_s,p_s)
      \, \delta_{T_t-T_s} \, ds}{\dps \int_0^t \delta_{T_t-T_s} \,
    ds} \right ) \, dt, \\
  \end{array}
  \right.
\end{equation}
where $W_t$ is still a standard $3N$-dimensional Brownian motion,
and $\delta$ is a Dirac mass.
The estimate of the update term is performed using a 
trajectorial estimate of the canonical average of $A$, 
using only the previous configurations sampled at the
same temperature. The difference between the single trajectory
discretization and the many replica implementation is illustrated in
Figure~\ref{fig:single_traj_discretization}.

The dynamics~(\ref{eq:single_replica_dynamics}) is discretized using a
splitting procedure, with the decomposition
\[
\left \{
  \begin{array}{cl}
    dq_t = & \dps M^{-1} p_t \, dt, \\ [5pt]
    dp_t = & \dps -\nabla V(q_t) \, dt - \xi M^{-1} p_t \, dt
    + \sqrt{2 \xi k_{\rm B} T_t} \, dW_t, \\ [5pt]
    dT_t = & \dps 0, \\
  \end{array}
  \right. \quad
  \left \{
  \begin{array}{cl}
    dq_t = & \dps 0, \\ [5pt]
    dp_t = & \dps 0, \\ [5pt]
    dT_t = & \dps -\gamma \left ( \frac{\dps \int_0^t A(q_s,p_s)
      \, \delta_{T_t-T_s} \, ds}{\dps \int_0^t \delta_{T_t-T_s} \, ds}
    \right ) \, dt. \\
  \end{array}
  \right.
\]
First, the dynamics at a fixed temperature
is integrated using some numerical scheme
of the general form $(q^{n+1},p^{n+1}) = \Phi_{\Delta t, T^n}(q^n,p^n)$,
where $(q^n,p^n)$ denotes an approximation of a realization of
$(q_{n \Delta t},p_{n \Delta t})$, and the current temperature $T^n$ is a parameter
of the numerical scheme. For instance, the BBK scheme~\cite{BBK84} with the modification
of~\cite{shardlow03} may be used for
Langevin dynamics ($i=1,\dots,N$ indexes the particles):
\[
\Phi_{\Delta t, T^n} \, : \, \left \{
\begin{array}{c @{= \ } l}
  p_{i}^{n+1/2} & p_{i}^{n} + \dps \frac{\Delta t}{2} \left ( - \nabla_{q_{i}}
    V (q^{n}) - \xi \frac{p_{i}^{n}}{m_{i}}
    + \frac12 \sqrt{2\xi\Delta t \, k_\mathrm{B} T^n} \, G_{i}^{n} \right ),\\[10pt]
  q_{i}^{n+1} & \dps q_{i}^{n} + \Delta t \
    \frac{p_{i}^{n+1/2}}{m_{i}}, \\ [10pt]
  p_{i}^{n+1} & \dps \left (1 + \frac{\xi \Delta t}{2 m_i} \right )^{-1}
    \ \left( p_{i}^{n+1/2} - \frac{\Delta t}{2} \nabla_{q_{i}} V
    (q^{n+1}) + \frac12 \sqrt{2\xi\Delta t \, k_\mathrm{B} T^n} \, G_{i}^{n} \right),\\
\end{array}
\right.
\]
where $\{G_i^n\}_{i,n}$ are independent standard $3$-dimensional Gaussian random vectors.
Of course, there many other implementations of the Langevin dynamics (see for
instance~\cite{SI02,WS03} and references in~\cite{CLS07}).
The temperature is then updated after
computing the conditional average of the observable $A$ along the
trajectory. Finally, the algorithm reads
\begin{equation}
  \label{eq:num_dyn}
  \left \{ \begin{array}{ccl}
    (q^{n+1},p^{n+1}) & = & \Phi_{\Delta t, T^n}(q^n,p^n), \\ [5pt]
    T^{n+1} & = & \dps T^n - \left( \frac{\dps \sum_{m=0}^{n} A(q^m,p^m)
      \, \chi_{\Delta T}(T^{m}-T^n)}{\dps \sum_{m=0}^{n} \chi_{\Delta T}(T^{m}-T^n)}
    \right ) \gamma \Delta t.\\
  \end{array} \right.
\end{equation}
The ratio in the update term of the temperature 
is always well-defined provided $\chi_{\Delta T} \geq 0$ 
and $\chi_{\Delta T}(0) > 0$.
The functions $\chi_{\Delta T}$ are approximations of Delta functions,
and correspond to some binning procedure.
For a given temperature, the corresponding
bin in the histogram is sought for, and the average value of the
observable for this temperature as well as the total number of visits
to this bin are updated. More precisely, discretizing the temperature
space into regions of width $\Delta T$, a given temperature $T$
can be rewritten as $T = \text{E}(T/\Delta T) \Delta T + R_T$, where
$\text{E}(x)$ denotes the closest integer to $x$ and $-1/2 < R_T
\leq 1/2$. Then,
\[
\chi_{\Delta T}(T^m-T^n) = \left \{ \begin{array}{cl}
    1/\Delta T & \text{when } \text{E}(T^m/\Delta T)=\text{E}(T^n/\Delta T), \\
    0 & \text{otherwise}.
\end{array} \right.
\]
The width $\Delta T$ controls the precision of the final result.
It should not be too small, so that the estimates of the
(trajectorial) conditional expectations are performed over enough
configurations.
It may however be refined during the simulation, once close enough to
the target temperature.

The choice of parameters, in particular the
parameters of the sampling dynamics (such as
Nos\'e-Hoover mass or Langevin friction) and $\gamma$ required
for the update of the temperature, is discussed in a specific case in
Section~\ref{sec:choice_parameters}.

\subsection{A test case: The computation of Hugoniot curves}
\label{sec:Hugoniot}

We present an application of the general sampling algorithm presented
in Section~\ref{sec:single_traj_discretization} to the Hugoniot
problem described in Section~\ref{sec:Hugoniot_pbm}, in the case of
noble gases such as Argon. This problem was already studied
in~\cite{MMSRLGH00}, where reference results
obtained from shock wave simulations are reported.

The interactions within noble gas atoms
are well-described by a Lennard-Jones potential:
\[
V(q_1,\dots,q_N) = \sum_{1 \leq i < j \leq N} v(|q_i-q_j|),
\qquad
v(r) = 4\varepsilon \left ( \left(\frac{\sigma}{r}\right)^{12}
-\left(\frac{\sigma}{r}\right)^{6} \right ).
\]
In the case of Argon, $\varepsilon/k_{\rm B} = 120$~K and $\sigma=3.405$~\AA.
It is convenient to use dimensionless parameters to present the
results, as is done in~\cite{MMSRLGH00}.
The reduced unit of distance $r_0 = 2^{1/6} \sigma$
corresponds to the minimum of $v$, the unit of mass and
energy are respectively the mass of an atom $m$ and
$\varepsilon$. Therefore, the unit of time is $\tau = r_0
\sqrt{m/\varepsilon}$.
For Argon, $r_0 = 3.82$~\AA, $m=6.64 \times 10^{-26}$~kg, $\varepsilon
= 1.66 \times 10^{-21}$~J, and $\tau = 2.42 \times 10^{-12}$~s.
The cut-off used for the computation of the forces is $R_{\rm cut} =
2.5 \sigma$, and we chose $\Delta T = 0.2$~K ($\Delta T = 1.67 \times 10^{-3}$
in reduced units) for the temperature histograms.

The results presented here are obtained for a crystal reference state,
in the cubic face centered geometry,
at $T_0 = 10$~K with initial density  $\rho_0 = 1.806 \times
10^3$~kg/m$^3$. The density is chosen so that the initial pressure
$P_0 \simeq 0$.

\subsubsection{Choice of parameters}
\label{sec:choice_parameters}

In the case of Hugoniot curves, it is convenient to rewrite the
parameter $\gamma$ arising in the update equation for the
temperature~(\ref{eq:single_replica_dynamics}) as
\[
\gamma = \frac{\nu}{N k_{\rm B}},
\]
where $\nu > 0$ is a frequency. The temperature update can then be
recast in a form involving only dimensionless quantities:
\[
d\left( \frac{T_t}{T_\text{ref}} \right )
= - \frac{\mathcal{A}_t(T_t)}{N k_{\rm B} T_{\rm ref}}  \, \nu \, dt,
\qquad
\mathcal{A}_t(T) = 
\frac{\dps \int_0^t A_c(q_s,p_s) \, \delta_{T-T_s} \, ds}
     {\dps \int_0^t \delta_{T-T_s} \, ds},
\]
since the observable $A_c$ (or $A_{xx,c}$), defined
by~(\ref{eq:function_hugoniot}) is an extensive quantity
homogeneous to an energy.
This is then also the case for the average quantity $\mathcal{A}_t(T)$.
The energy $E_\text{ref} = k_{\rm B} T_\text{ref}$ is
some reference energy per particle.
We first discuss the choice of the reference
temperature, and then the choice of the frequency $\nu$.

The other parameters have standard values in molecular dynamics
simulations: the time step $\Delta t \simeq 0.001$ in reduced
units, while the Langevin friction $\xi$ is such that $\xi/m \sim
1/\tau$. Our simulations for Argon were performed with $\Delta t =
2 \times 10^{-15}$~s$^{-1}$ and $\xi/m = 10^{12}$~s$^{-1}$.

\paragraph{Reference temperature.}
\label{sec:reference_temperature}

The reference temperature is computed from the initial configuration
using the estimator of the temperature derived
in~\cite{MMSRLGH00}. Decomposing the microscopic observables
associated with the energy and the pressure as
\[
H = H_\text{kin} + H_\text{pot}, \qquad P = P_\text{kin} + P_\text{pot},
\]
with
\[
H_\text{kin} = \frac12 p^T M^{-1} p, \qquad
P_\text{kin} = \frac{1}{3 |\mathcal{D}|} p^T M^{-1} p,
\]
the Hugoniot problem~(\ref{eq:Hugoniostat_problem}) implies for a
given compression~$c$:
\[
T = \frac{1}{N k_{\rm B}} \frac{2}{4-c} \left ( \langle H \rangle_{|\mathcal{D}|_0,T_0} -
\langle V \rangle_{c|\mathcal{D}_0|,T} + \frac12 (
\langle P_\text{pot} \rangle_{c\mathcal{D}_0,T} + \langle P \rangle_{|\mathcal{D}|_0,T_0})
(1-c)|\mathcal{D}_0|  \right ).
\]
Therefore, an estimator of the temperature in terms of the positions
$q$ of the particles only is
\begin{equation}
  \label{eq:estimatorT}
  \widehat{T}(q) = \frac{1}{N k_{\rm B}} \frac{2}{4-c}
  \left ( \langle H \rangle_{|\mathcal{D}_0|,T_0} - V(q)
  + \frac12 (P_\text{pot}(q)+ \langle P \rangle_{|\mathcal{D}_0|,T_0} )
  (1-c)|\mathcal{D}_0| \right ),
\end{equation}
which is by construction such that
$\dps \left \langle \widehat{T} \right \rangle_{|\mathcal{D}|,T} = T$.
The reference temperature is chosen to be
\begin{equation}
  \label{eq:reference_temperature}
  T_{\rm ref} = \widehat{T}(q^0),
\end{equation}
where $q^0$ denotes the initial configuration. Usually, this initial
configuration is obtained by compressing a perfect lattice in the
chosen direction(s). When the compression is not isotropic, the
estimator~(\ref{eq:estimatorT}) for the initial temperature should be
replaced by a similar estimator where the isotropic pressure
observable $P$ is replaced by the observable associated with the
$P_{xx}$ component of the pressure tensor. The decomposition into
kinetic and potential parts of $P_{xx}$ is done as for $P$.

\paragraph{Typical frequency.}
\label{sec:frequency}

\begin{figure}[h]
\center
\includegraphics[width=10cm]{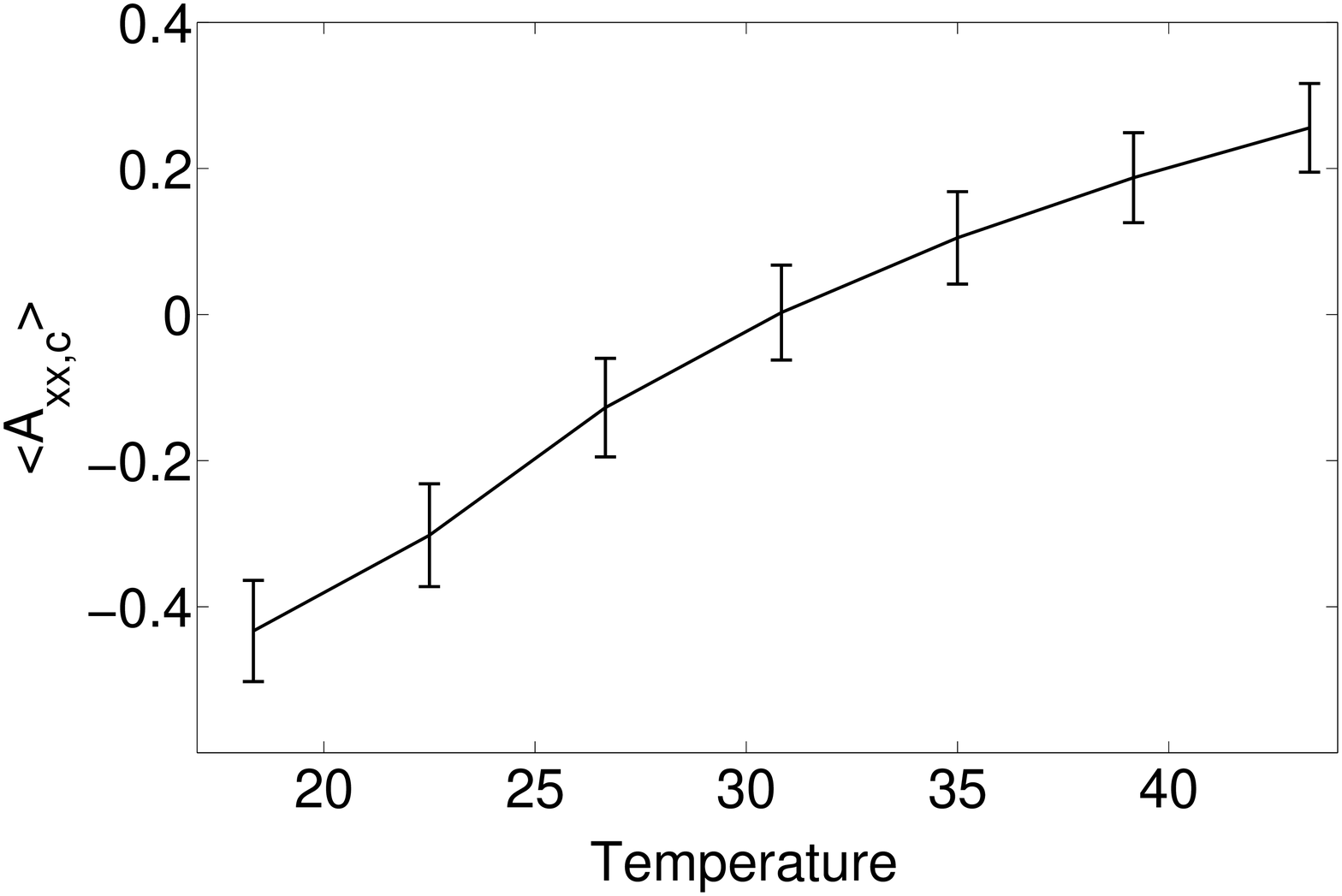}
\caption{\label{fig:average_Ac}
  Distribution of $\overline{A}_{xx,c}$ (in reduced units)
  for an anisotropic compression $c=0.62$ as a function of
  temperature. The error bars are equal to twice the standard
  deviation of $\overline{A}_{xx,c}$ obtained for a system of $N=4,000$
  atoms, and are meant to give an order of magnitude of the width of the
  distribution of values of $\overline{A}_{xx,c}$.
}
\end{figure}

Once the reference temperature is set, it is possible
to estimate the typical magnitude of the normalized observable
\[
\overline{A}_c(q,p) = \frac{A_c(q,p)}{N k_\text{B} T_\text{ref}},
\]
for instance by performing some short canonical samplings at fixed
compressions and temperatures. The quantity $\overline{A}_{xx,c}$ is
defined and estimated in a similar manner.
In the case of Hugoniot sampling, Figure~\ref{fig:average_Ac}
presents the distribution of $\overline{A}_{xx,c}$ for canonical samplings
performed at different temperatures and anisotropic compressions.
The typical values of $\overline{A}_{xx,c}$
are of order $\triangle \overline{A}_{xx,c} \sim 0.5$. This
gives a typical size for the term appearing in the
equation on the temperature since 
the temperature is updated by a factor of order
$\triangle \overline{A}_{xx,c} \nu \Delta t$ at each time step.
The magnitude of $\nu$ should be chosen such
that the evolution of the temperature is not too
fast (as suggested by Theorems~\ref{thm:LSIcv} and~\ref{thm:Poincare_cv}). For instance,
\begin{equation}
  \label{eq:rule_nu}
  \frac{\delta T}{T_\text{ref}} = \triangle \overline{A}_{xx,c} \nu
    \Delta t \sim 10^{-3} - 10^{-1}.
\end{equation}
Typical time integration steps are $\Delta t = 10^{-3} -
10^{-2}$ in reduced units. The parameter $\nu$ can then be found from
the above heuristic rule. 

\begin{figure}[h]
\center
\includegraphics[width=7.3cm]{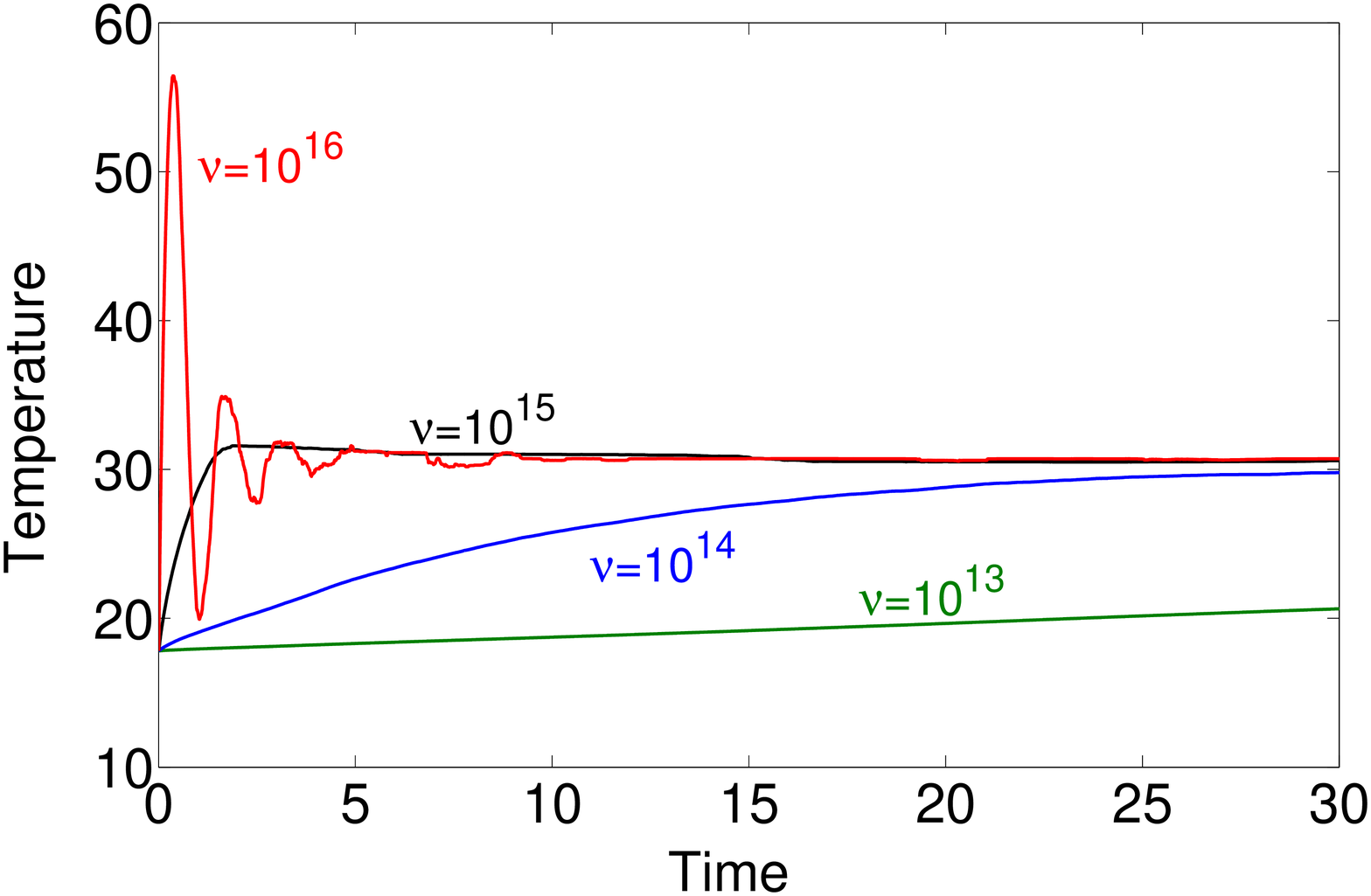}\hfill
\includegraphics[width=7.3cm]{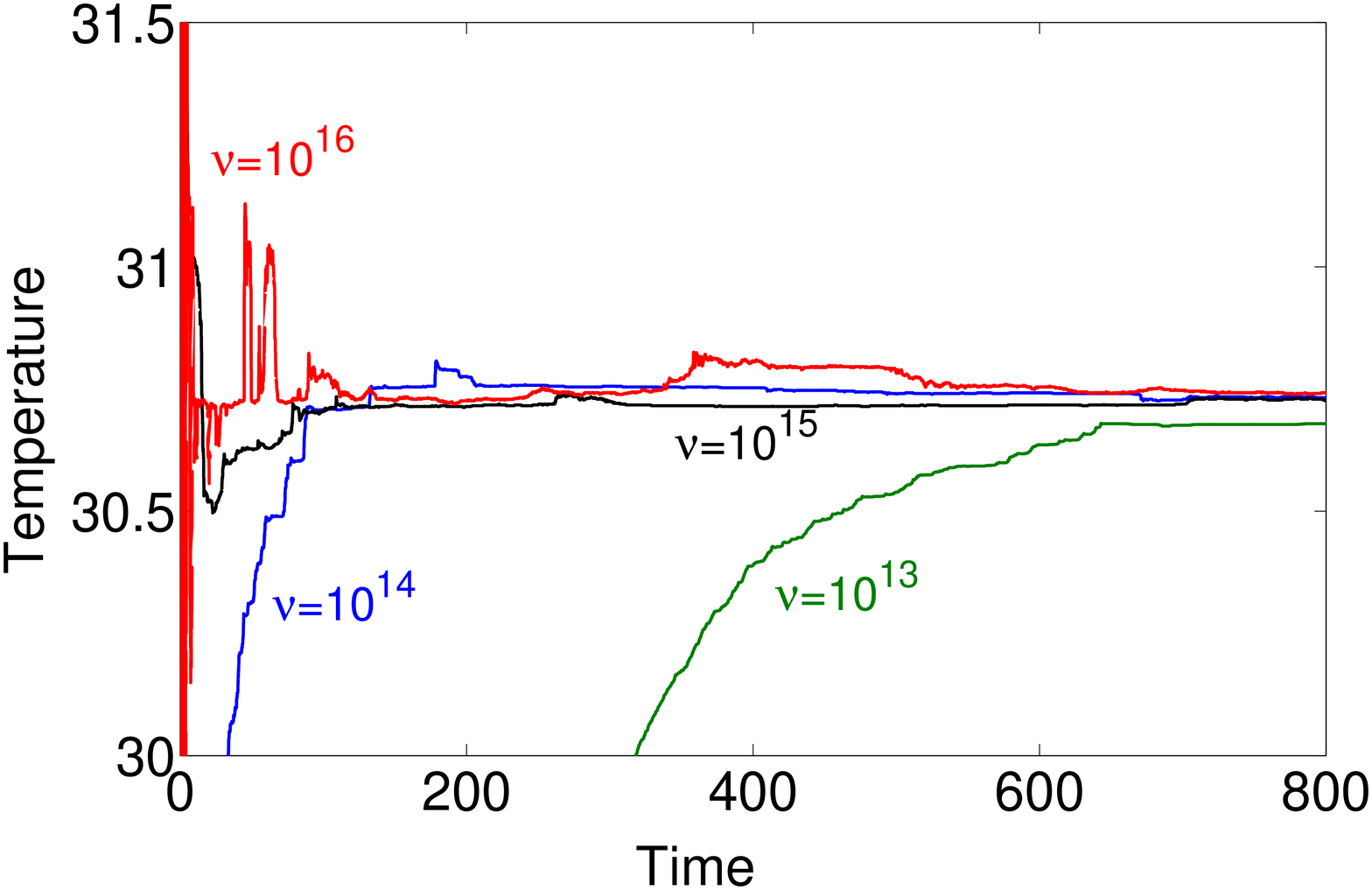}
\caption{\label{fig:nu}
  (color online)
  Plot of the temperature as a function of time (in reduced units) for
  different values of the frequency $\nu$ (in $s^{-1}$), for a system of size $N=4,000$.
}
\end{figure}

\begin{figure}[h]
\center
\includegraphics[width=7.3cm]{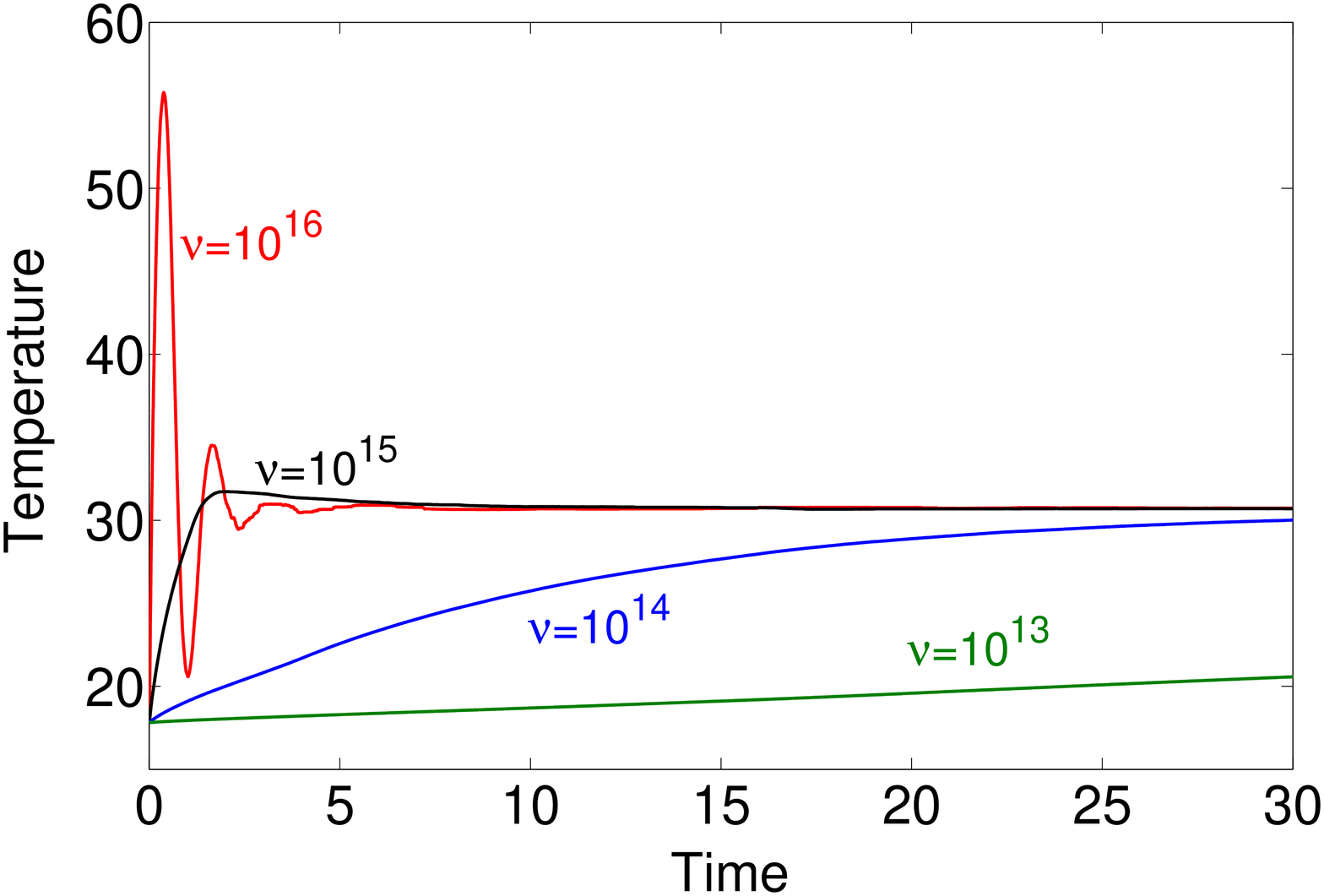}\hfill
\includegraphics[width=7.3cm]{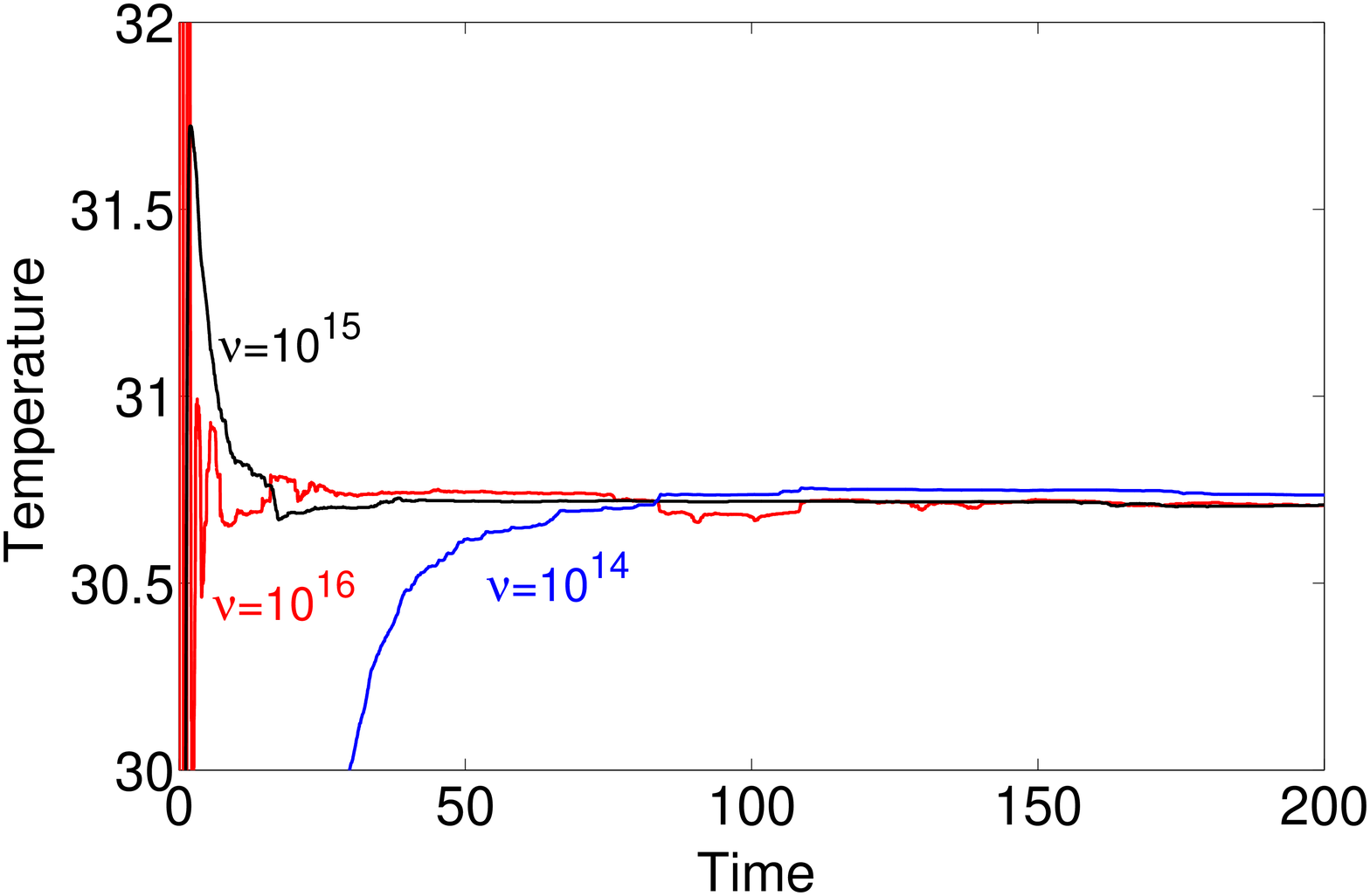}
\caption{\label{fig:nu2}
  (color online) 
  Plot of the temperature as a function of time (in reduced units) for
  different values of the frequency $\nu$ (in $s^{-1}$), for a system of size $N=32,000$.
}
\end{figure}

In the Hugoniot example, the rule~(\ref{eq:rule_nu}) gives an estimate
of possible values of the frequency.
For Argon at a compression $c=0.62$, $\nu=10^{15}$~s$^{-1}$ seems a reasonable choice.
The results of Figures~\ref{fig:nu} and~\ref{fig:nu2}, obtained
for systems of $N=4,000$ and $N=32,000$ atoms respectively, show that the temperature
converges with the method presented here, and that frequencies around
$\nu=10^{15}$~s$^{-1}$ are indeed a reasonable choice.
For smaller frequencies, the convergence is slower, whereas larger
frequencies trigger fast initial oscillations which may lead to
numerical instabilities in the scheme.
In any case however, the final result does not
depend on the value of $\nu$, which shows the robustness of the method.
It may be interesting to increase the value of $\nu$ as the typical
values of the observable decrease during the simulation (see also Remark~\ref{rmk:extensions}).

%
%

\subsubsection{Hugoniot curve}

\begin{figure}[h]
\center
\includegraphics[width=10cm]{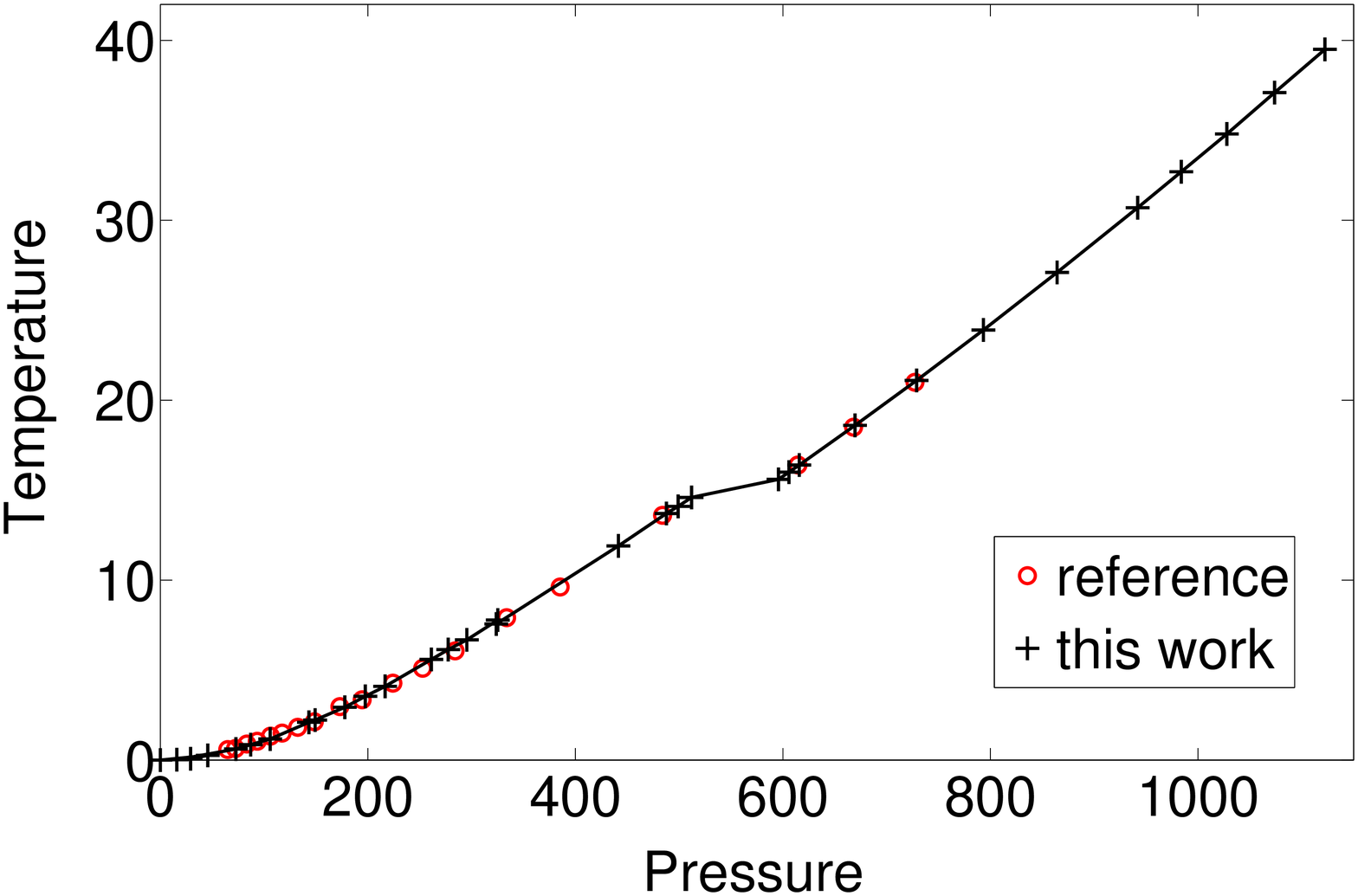}
\caption{\label{fig:hug}
  (color online) 
  Hugoniot curve for Argon (in reduced units), computed using the
  method presented in this work (black solid line and crosses). The 
  reference results from \cite{MMSRLGH00} are also reported (red circles).
}
\end{figure}

We present in Figure~\ref{fig:hug} the Hugoniot curve
obtained for a system of $N=4,000$ particles,
using the parameters given in
Section~\ref{sec:choice_parameters}, with $\nu = 10^{14}$~s$^{-1}$
for compressions $c \leq 0.7$ (temperatures
around 4 and less, in reduced units) and $\nu = 10^{15}$~s$^{-1}$ for
higher compressions.
This curve is obtained by considering many different anisotropic compressions,
and computing the temperature as well as the associated average pressure
$P$ in the system.
The results are in very good agreement with~\cite{MMSRLGH00}, which validates the method.

\section{Proofs of the mathematical results}
\label{sec:proofs}

\subsection{Proof of Theorem~\ref{thm:ExistenceUniqueness}}
\label{sec:proof_existence}

\paragraph{Existence of a time $\tau > 0$ 
  where the temperature remains positive.}

Assume that $\psi(t,\cdot) \in \mathrm{L}^1(\mathcal{M})$ and $\psi
\geq 0$, so that $\| \psi(t,\cdot) \|_{\mathrm{L}^1} = 1$. Then,
since the temperature derivative is bounded by $\gamma
\| A \|_\infty$,
there exists a time $\tau > 0$ such that $|T(t)-T^0| \leq T^0/2$ for all $t
\in [0,\tau]$. For instance, 
\begin{equation}
  \label{eq:choice_tau}
  \tau = \frac{T^0}{2\gamma \| A \|_\infty} > 0,
\end{equation}
where $\| A \|_\infty = \sup \, \{ \, |A(q)|, \ q \in \mathcal{M} \, \}$. 
This choice ensures that the temperature remains positive.
We will show below that $\psi \in \mathrm{L}^1(\mathcal{M})$ since 
$\psi \in \mathrm{L}^2(\mathcal{M})$ and $|\mathcal{M}| < +\infty$.

\paragraph{Construction of a solution through a fixed point application.}
 
We show the existence of a solution using a fixed-point theorem.
Denote 
\[
E = \mC^1([0,\tau],\mathbb{R}) \cap \left \{ \frac{T^0}{2} \leq
T \leq \frac{3T^0}{2} \right \} \cap \{ T(0) = T^0 \} \cap \{ |T'|
\leq \gamma \| A \|_\infty\}.
\]
This set is a bounded closed convex subset of $\mC^1([0,\tau],\mathbb{R})$.
Consider the mapping
\begin{equation}
  \label{eq:g_function}
  g \ : \ \begin{array}{ccccccc}
  E  & \longrightarrow &
  \mC^0([0,\tau],\mathrm{H}^2(\mathcal{M})) 
  & \longrightarrow & \mC^2([0,\tau],\mathbb{R}) \cap E
  & \longrightarrow & E \\
  T & \longmapsto & \psi_T & \longmapsto & g(T) & \longmapsto & g(T) \\ 
  \end{array} 
\end{equation}
The construction of $\psi_T$ from a given temperature function $T$ is performed
using the equation 
\begin{equation}
  \label{eq:psi_given_T}
  \partial_t \psi_T = k_{\rm B} T(t) \, \nabla \cdot \left [ \mu_{T(t)}
    \nabla \left ( \frac{\psi_T}{\mu_{T(t)}}\right ) \right ]
  = k_{\rm B} T(t) \Delta \psi_T + \nabla V \cdot \nabla \psi_T + \Delta V \psi_T,
\end{equation}
and the temperature function $g(T)$ is then recovered using the second equation
in \eqref{eq:adaptive_dynamics_PDE}, namely
\[
g(T)(t) = T^0 - \gamma \int_0^t \int_\mathcal{M} A(q) \, \psi_T(s,q)
\, dq \, ds.
\] 

\paragraph{Regularity of $\psi_T$ given $T$.}

The equation \eqref{eq:psi_given_T} is of the general type 
\begin{equation}
  \label{eq:KatoFormulation}
  \partial_t \psi_T + \mathcal{L}_T(t) \psi_T = 0, 
\end{equation}
where the family of operators $\{ \mathcal{L}_T(t) \}_{t \geq 0}$ 
have constant domains $D = \mathrm{H}^2(\mathcal{M})$, and 
\[
\mathcal{L}_T(t) \psi = - k_{\rm B} T(t) \, \nabla \cdot \left [ \mu_{T(t)}
  \nabla \left ( \frac{\psi}{\mu_{T(t)}}\right ) \right ] = -\nabla
  \cdot \left ( k_{\rm
  B}T(t) \, \nabla \psi + \nabla V \psi \right ).
\]
It is enough to show the regularity of the solution of 
$\partial_t \widetilde{\psi} + (\eta  + \mathcal{L}_T(t))
\widetilde{\psi} = 0$ since the latter equation is related to 
the original equation \eqref{eq:KatoFormulation} through the transform
$\psi_T = \exp(\eta t) \widetilde{\psi}$.

For a fixed $t$ and $f \in \mathrm{L}^2(\mathcal{M})$, consider the equation 
\begin{equation}
  \label{eq:LaxMilgram}
  (\eta  + \mathcal{L}_T(t)) u = f.
\end{equation}
Lax-Milgram's theorem shows that there is a unique solution $u \in
\mathrm{H}^1(\mathcal{M})$. Indeed, 
$(u,v) \mapsto \langle u, (\eta + \mathcal{L}_T(t))v \rangle$ can be
extended to a functional on $\mathrm{H}^1(\mathcal{M})^2$ using
\[
\left | \langle u, (\eta + \mathcal{L}_T(t))v \rangle \right |
= \left | \int_\mathcal{M} \eta u v + k_{\rm B} T(t) \nabla u \cdot
\nabla v + v \nabla u \cdot \nabla V\right |  \\
\leq C \| u \|_{\mathrm{H}^1} \, \| v \|_{\mathrm{H}^1}, \\
\]
where the constant $C$ depends only on $\| \nabla V \|_\infty$ and 
on $T^0$ (actually on the upper bound of the temperature).
Besides, 
$u \in D \mapsto  \langle u, (\eta + \mathcal{L}_T(t))u \rangle $, 
extended to a quadratic form on $\mathrm{H}^1(\mathcal{M})$, is
coercive on this space for $\eta$ large enough since
\begin{eqnarray*}
\langle u, (\eta + \mathcal{L}_T(t))u \rangle 
& = & \int_\mathcal{M} \eta \, u^2 + k_{\rm B} T(t) |\nabla u|^2 
+ u \nabla u \cdot \nabla V \\
& \geq & \frac{k_{\rm B} T^0}{2} \int_\mathcal{M} |\nabla u|^2
+ \eta \int_\mathcal{M} u^2 - \frac12 \| \nabla V\|_\infty \left (
\alpha \int_\mathcal{M} |\nabla u|^2 + \frac1\alpha \int_\mathcal{M}
u^2 \right ) \\
& \geq & c \| u \|_{\mathrm{H}^1}^2, \\
\end{eqnarray*}
for $\eta$ large enough (choosing first $\alpha < k_{\rm B} T^0/\|
\nabla V\|_\infty$ and then $\eta > \| \nabla V \|_\infty/2\alpha$).
Again, the constant $c$ depends only on $\| \nabla V \|_\infty$ and on
$T^0$. The solution $u$ of \eqref{eq:LaxMilgram}
is in fact such that 
\[
\Delta u = \frac{1}{k_{\rm B}T(t)} \left ( \eta u - \nabla V \cdot
\nabla u - u \Delta V - f\right ) \in \mathrm{L}^2(\mathcal{M}),
\]
so that $u \in D$.
It is easily seen that 
the mapping $f \mapsto u = (\lambda + \eta + \mathcal{L}_T(t))^{-1}f$ for $\lambda
\in \mathbb{C}$ with $\mathrm{Re}(\lambda)
\geq 0$ is continous, 
and bounded by $M/(1+|\lambda|)$ for some $M > 0$.

Besides, for $0 \leq r,s,t \leq \tau$, 
\begin{eqnarray*}
  \| \left [ (\eta + \mathcal{L}_T(r)) - (\eta + \mathcal{L}_T(s))
  \right ] (\eta + \mathcal{L}_T(t))^{-1}
  \|_{\mathcal{L}(\mathrm{L}^2,\mathrm{L}^2)} & = & 
  k_{\rm B} |T(r)-T(s)|
  \| \Delta (\eta + \mathcal{L}_T(t))^{-1}
  \|_{\mathcal{L}(\mathrm{L}^2,\mathrm{L}^2)} \\
  & \leq & c |T(r)-T(s)| \leq C |r-s|,
\end{eqnarray*}
for some constants $C,c > 0$ since $T$ is $\mathrm{C}^1$.

The results of Kato \cite{Kato61} then 
show the existence of a unique solution 
$\psi_T \in \mathrm{C}^0([0,\tau],\mathrm{H}^2(\mathcal{M}))$ and hence 
$\psi_T \in \mathrm{C}^0([0,\tau],\mathrm{H}^1(\mathcal{M}))$.
Bounds on the $\mathrm{H}^1(\mathcal{M})$ norm of $\psi_T$ in terms of
the bounds on $T$ are obtained with a Gronwall inequality:
\begin{eqnarray*}
\partial_t \left ( \frac12 \| \psi_T \|^2_{\mathrm{H}^1} \right )
& = & \int_\mathcal{M} (\psi_T-\Delta \psi_T) 
\nabla \left ( k_{\rm B} T(t) \nabla \psi_T + \nabla V \psi_T \right ) \\
& = & -k_{\rm B} T(t) \int_\mathcal{M} |\nabla \psi_T|^2 + |\Delta
\psi_T|^2 - \int_\mathcal{M} \Delta \psi_T \nabla V \cdot \nabla
\psi_T \\
& & \quad + \int_\mathcal{M} |\psi_T|^2 \Delta V + \psi_T \nabla V \cdot
\nabla \psi_T - \int_\mathcal{M} \psi_T \Delta \psi_T \Delta V. \\
\end{eqnarray*}
Using the following Cauchy-Schwarz inequalities to bound the terms
involving $\Delta \psi_T$: 
\[
\left | \int_\mathcal{M} \Delta \psi_T \nabla V \cdot \nabla
\psi_T \right | \leq \frac{k_{\rm B}T^0}{4} \int_\mathcal{M} |\Delta \psi_T|^2 + 
\frac{\| \nabla V\|^2_\infty}{k_{\rm B}T^0} \int_\mathcal{M} |\nabla \psi_T|^2,
\]
\[
\left | \int_\mathcal{M} \psi_T \Delta \psi_T \Delta V \right | \leq
\frac{k_{\rm B}T^0}{4} \int_\mathcal{M} |\Delta \psi_T|^2 + 
\frac{\| \Delta V\|^2_\infty}{k_{\rm B}T^0} \int_\mathcal{M} |\psi_T|^2,
\]
it follows
\begin{eqnarray*}
\partial_t \left ( \frac12 \| \psi_T \|^2_{\mathrm{H}^1} \right )
& \leq & -k_{\rm B} T(t) \int_\mathcal{M} |\nabla \psi_T|^2 
+ \int_\mathcal{M} |\psi_T|^2 \Delta V + \psi_T \nabla V \cdot
\nabla \psi_T + \frac{\| \nabla V\|^2_\infty + \| \Delta V\|^2_\infty}{k_{\rm B}T^0}
\| \psi_T \|^2_{\mathrm{H}^1} \\
& \leq & K \| \psi_T \|^2_{\mathrm{H}^1},
\end{eqnarray*}
where the constant $K > 0$ depends only on $V$ and its
derivatives, and on $T^0$ (actually, the lower bound of the temperature). This shows that 
\[
\| \psi_T(t,\cdot) \|_{\mathrm{H}^1} \leq \exp(\tau K) \| \psi^0 \|_{\mathrm{H}^1}
\]
for all $0 \leq t \leq \tau$.

\paragraph{Regularity of $g(T)$ given $\psi$.}

From the equation on the temperature:
\[
g(T)'(t) = -\gamma \int_\mathcal{M} A(q) \, \psi_T(t,q) \, dq,
\]
it follows
\[
g(T)''(t) = \gamma \int_\mathcal{M} (\nabla A \cdot \nabla V -
k_{\rm B}T(t) \Delta A) \psi_T,
\]
and 
\[
g(T)^{(3)}(t) = -\gamma \int_\mathcal{M} \nabla \cdot (\nabla A \cdot \nabla V -
k_{\rm B}T(t) \Delta A) (\nabla V \psi_T + k_{\rm B}T(t) \nabla \psi_T)
- \gamma k_{\rm B}T'(t) \int_\mathcal{M} \Delta A \psi_T.
\]
Therefore, $|g(T)^{(3)}(t)| \leq \gamma C_{A,V} 
\| T\|_{\mathrm{C}^1} (1 + \| T\|_{\mathrm{C}^1}) 
\,\| \psi_T(t,\cdot) \|_{\mathrm{H}^1}$, where
$C_{A,V}$ depends only on $\|\mathcal{M}|$ and on $\mathrm{L}^\infty$ bounds on $A,V$ and
their derivatives (up to the order 3 for $A$, and to the order 2 for $V$).
Uniform bounds on $g(T)''$ are recovered
by time integration:
\begin{eqnarray*}
\sup_{0 \leq t \leq \tau} | g(T)''(t)| 
& \leq & |g(T)''(0)| + \tau \| g(T)^{(3)}\|_\infty \\
& \leq & \gamma \left (\widetilde{C}_{A,V} \| \psi^0 \|_{\mathrm{H}^1} 
+ \tau C_{A,V} \| T\|_{\mathrm{C}^1} \, (1 + \| T\|_{\mathrm{C}^1})
\| \psi \|_{\mathrm{C}^0([0,T],\mathrm{H}^1)} \right ) \\
& \leq & \gamma \left [ \widetilde{C}_{A,V} + \tau C_{A,V} \|
T\|_{\mathrm{C}^1} (1 + \| T\|_{\mathrm{C}^1}) \, \exp(K \tau) \right ] \| \psi^0 \|_{\mathrm{H}^1}
\leq M,
\end{eqnarray*}
for some $M > 0$ (depending on $\tau$, $\psi^0$, $T^0$).
Since $\|g(T)'\|_\infty \leq \gamma \| A \|_\infty$, bounds on $g(T),
g(T)'$ consistent with the set $E$ are still valid with the choice of
$\tau$ given by \eqref{eq:choice_tau}.
This shows finally that $g(T)$ belongs to a bounded 
set of $\mC^2([0,\tau],\mathbb{R})$.

\paragraph{Existence of a fixed point.}

The mapping $g$ is continuous from the
bounded closed convex set $E$
to istself, and in fact compact since the injection 
$\mC^2([0,\tau],\mathbb{R}) \cap \{ |T''| \leq M \} \cap E \hookrightarrow E$ is compact. 
Schauder's theorem (see for instance \cite{Zeidler}) shows that
there exists $T \in E$ such that $g(T) = T$, therefore giving the
existence of a solution to the equation \eqref{eq:adaptive_dynamics_PDE}.

\paragraph{Uniqueness.}

Consider two solutions $(T_1,\psi_1)$ and $(T_2,\psi_2)$ starting from
the same initial condition $(T^0,\psi^0)$. Then, on the interval
$[0,\tau]$, 
\begin{eqnarray*}
  \frac{d}{dt} \left ( \frac12 (T_1(t)-T_2(t))^2 \right )
  & = & - \gamma (T_1(t)-T_2(t)) \int_\mathcal{M} A \, (\psi_1-\psi_2) \\
  & \leq & \sqrt{|\mathcal{M}|}\| A \|_\infty  \, |T_1(t)-T_2(t)| \,
  \| \psi_1 - \psi_2 \|_{\mathrm{L}^2}.
\end{eqnarray*}
Moreover, 
\begin{eqnarray*}
  \frac{d}{dt} \left ( \frac12 \| \psi_1 - \psi_2 \|_{\mathrm{L}^2}^2 \right )
  & = & - \int_\mathcal{M} \nabla V \cdot \nabla (\psi_1-\psi_2)
  (\psi_1 - \psi_2) - k_{\rm B} \frac{T_1(t)+T_2(t)}{2} \int_\mathcal{M} |\nabla
  (\psi_1-\psi_2)|^2 \\
  & & \quad - k_{\rm B} \frac{T_1(t)-T_2(t)}{2} \int_\mathcal{M} \nabla
  (\psi_1 + \psi_2) \cdot \nabla (\psi_1 - \psi_2) \\
  & \leq & \quad c_1 \| \psi_1 - \psi_2 \|_{\mathrm{L}^2}^2 + c_2 |T_1(t)-T_2(t)|^2
  \int_\mathcal{M} |\nabla (\psi_1+\psi_2)|^2, \\
\end{eqnarray*}
for some constants $c_1,c_2 >0$. Indeed, the
the $\nabla (\psi_1 - \psi_2)$ 
contributions of the first and third terms on the right hand side of
the first line can be cancelled
using the inequality $2ab \leq a^2/\eta + \eta
b^2$ and
\[
- k_{\rm B} \frac{T_1(t)+T_2(t)}{2} \int_\mathcal{M} |\nabla
(\psi_1-\psi_2)|^2 \leq -\frac{k_{\rm B} T^0}{2} \int_\mathcal{M} |\nabla
(\psi_1-\psi_2)|^2.
\]
In view of the uniform bounds on $\|\nabla \psi_1(t,\cdot)\|_{\mathrm{L}^2}$, $\|\nabla
\psi_2(t,\cdot)\|_{\mathrm{L}^2}$,
\[
\frac{d}{dt} \left ( \frac12 (T_1(t)-T_2(t))^2 + 
\frac12 \| \psi_1 - \psi_2 \|_{\mathrm{L}^2}^2 \right ) \leq C 
\left ( \frac12 (T_1(t)-T_2(t))^2 + 
\frac12 \| \psi_1 - \psi_2 \|_{\mathrm{L}^2}^2 \right )
\]
for some $C > 0$. As
\[
\frac12 (T_1(0)-T_2(0))^2 + 
\frac12 \| \psi_1(0,\cdot) - \psi_2(0,\cdot) \|_{\mathrm{L}^2}^2 = 0,
\]
a Gronwall inequality shows that $T_1 = T_2$ and $\psi_1 = \psi_2$ for
all $0 \leq t \leq \tau$.

\paragraph{Positivity of $\psi$ for relative entropy estimates.}

It is important that $\psi > 0$ for the Fisher information to be
defined in the case of relative entropy estimates (see the proof of
Theorem~\ref{thm:LSIcv} in Section~\ref{sec:proof_LSIcv}). In any cases, the
density $\psi$ has to remain non-negative.

Consider
\begin{equation}
  \label{eq:ChangeOfTime}
  \Phi(t,q) = \psi(s(t),q), 
\end{equation}
where the function $t \mapsto s(t)$ is defined through the
ordinary differential equation: 
\[
s'(t) = \frac{1}{k_{\rm B}T(s(t))}.
\]
The function $s$ is well defined (since $T > 0$), continuous and in
fact $\mathrm{C}^1$, and strictly increasing hence invertible.
The evolution of the density $\Phi$ is obtained by solving
\begin{equation}
  \label{eq:rescaled_eq}
  \partial_t \Phi = \Delta \Phi + \frac{1}{k_{\rm B} T(s(t)) } \nabla
  \cdot (\Phi \nabla V) = \Delta \Phi + \frac{1}{k_{\rm B} \widetilde{T}(t) } \nabla
  \cdot (\Phi \nabla V),
\end{equation}
with $\widetilde{T}(t) = T(s(t))$. The density $\psi$ is recovered with
\eqref{eq:ChangeOfTime} since $s$ is invertible.
The function $\phi = \Phi/\sqrt{\mu_{\widetilde{T}(t)}}$ satisfies
\[
\partial_t \phi(t,x) = \Delta \phi(t,x) + W(t,x) \phi(x),
\]
where the potential 
\[
W(t,x) = \frac{1}{2k_{\rm B}\widetilde{T}(t)} \left ( \Delta V(x)
-\frac{1}{2k_{\rm B}\widetilde{T}(t)} |\nabla V(x)|^2 - \frac{\widetilde{T}'(t)}{\widetilde{T}(t)} V(x) \right )
\]
is bounded.
Since $\mu_{T}$ is positive and uniformly bounded from below, it is enough to
show that $\phi(t,\cdot) > 0$ to have $\psi(t,\cdot) > 0$.
The function $\widetilde{\phi} = \exp(t\|W\|_\infty) \phi$ satisfies
\[
(\partial_t - \Delta) \widetilde{\phi} = (W(t,x) + \| W \|_\infty)
\widetilde{\phi} \geq 0.
\]
Therefore, $\widetilde{\phi}(t,\cdot) \geq \inf
\widetilde{\phi}(0,\cdot)$, and so, $\phi(t,x) \geq m
\exp(-t\|W\|_\infty) > 0$ provided 
$\phi(0,\cdot) \geq m > 0$. This shows that $\psi > 0$ provided
$\psi^0 > 0$. Similarly, $\psi \geq 0$ provided $\psi^0 \geq 0$.

\subsection{Proof of Theorem~\ref{thm:LSIcv}}
\label{sec:proof_LSIcv}

\paragraph{General strategy of the proof.}
We first consider the solution of \eqref{eq:adaptive_dynamics_PDE} 
on a time interval $[0,\tau]$ as given
by Theorem~\ref{thm:ExistenceUniqueness}, possibly reducing $\tau$ so
that $T(t) \in I_T^A \cap T_T^\mathrm{LSI}$ for all $0 \leq t \leq \tau$.
We then show (see below) that
this implies $\mathcal{E}'(t) \leq -\kappa \mathcal{E}(t)$ for some
some $\kappa > 0$. This will show that $\mathcal{E}$ decreases, and so
$T$ remains in $I_T^A \cap T_T^\mathrm{LSI}$ since $|T(t)-T^\ast| \leq
\sqrt{2\mathcal{E}(t)} \leq \sqrt{2\mathcal{E}(0)} \leq \sqrt{2\mathcal{E}^*}$.
The solution can then be continued at all times $t \geq 0$ using
repeatedly the arguments of the proof of
Theorem~\ref{thm:ExistenceUniqueness}, upon replacing the set $E$ in
the definition of 
$g$ given in \eqref{eq:g_function} by the closed convex bounded set
$E = \mathrm{C}^1([n\tau,(n+1)\tau],\mathbb{R}) \cap \{ T \in I_T^A \cap
T_T^\mathrm{LSI} \} \cap \{ T(n\tau) = T^{n\tau} \} \cap \{ |T'| \leq \gamma \|
A \|_\infty \}$, where $T^{n\tau}$ is the final value of the solution
on the interval $[(n-1)\tau,n\tau]$ (for $n \geq 1$). The main difference with the proof
of Section~\ref{sec:proof_existence} is that uniform 
(upper and lower) bounds on the temperature are provided thanks to the
decrease of the entropy.

\paragraph{Proof of the Gronwall inequality $\mathcal{E}'(t) \leq -\kappa \mathcal{E}(t)$.}
With the function $h$ chosen here,
\begin{eqnarray*}
\frac{dE(t)}{dt} & = & -k_{\rm B} T(t) \, 
\int_\mathcal{M} \frac{|\nabla f|^2}{f} \, \mu_{T(t)} 
- \frac{T'(t)}{k_\mathrm{B} T(t)^2}
\int_{\mathcal{M}} (f - 1)\, \left(
V - \left \langle V \right \rangle_{T(t)} \right ) \, \mu_{T(t)}\\
& \leq & -\rho k_{\rm B} T(t) \, E(t) + \frac{2 |T'(t)| \, \| V \|_\infty}{k_\mathrm{B} T(t)^2}
\int_{\mathcal{M}} | f - 1 | \, \mu_{T(t)}\\
& \leq & -\rho k_{\rm B} T(t) \, E(t) + \frac{4 |T'(t)| \, \| V \|_\infty}{k_\mathrm{B} T(t)^2}
\sqrt{E(t)}, \\
\end{eqnarray*}
where we have used the Csiz\'ar-Kullback inequality 
\[
\int_{\mathcal{M}} | f - 1 | \, \mu_{T(t)} \leq 2 \sqrt{E(t)}
\]
in the last step. 
The derivative of the temperature can be bounded using
\[
T'(t) = -\gamma \int_{\mathcal{M}} A \psi_t = -\gamma \left ( \left \langle A
\right \rangle_{T(t)} + \int_{\mathcal{M}} A (f - 1) \mu_{T(t)}\right ),
\]
which shows that 
\begin{eqnarray*}
|T'(t)| & \leq & \gamma \left ( \left| \left \langle A
\right \rangle_{T(t)} - \left \langle A
\right \rangle_{T^\ast}\right| + \left | \int_{\mathcal{M}} A \psi_t -
\int_{\mathcal{M}} A \mu_{T(t)} \right | \right ) \\
& \leq & \gamma \left ( a \left| T(t) - T^\ast \right| 
+ \left | \int_{\mathcal{M}} A \psi_t -
\int_{\mathcal{M}} A \mu_{T(t)} \right |\right) \\
& \leq & \gamma \left ( a \left| T(t) - T^\ast \right|  + 2 \| A
\|_\infty \sqrt{E(t)} \right ), \\
\end{eqnarray*}
using Assumption~\ref{ass:global_assumption} and 
the Csiz\'ar-Kullback inequality to bound the second term on the
right-hand side.
These estimates are also helpful for bounding the derivative of the
temperature entropy:
\begin{eqnarray*}
\frac{d}{dt} \left [ \frac12 (T(t)-T^\ast)^2 \right ] & = & 
-\gamma \left ( \left \langle A
\right \rangle_{T(t)} + \int_{\mathcal{M}} A (f - 1)
\mu_{T(t)}\right ) (T(t)-T^\ast)\\ 
& \leq & -\gamma \left ( \left \langle A
\right \rangle_{T(t)} - \left \langle A
\right \rangle_{T^\ast}\right ) (T(t)-T^\ast) + 
2 \gamma \| A \|_\infty \left| T(t) - T^\ast \right| \sqrt{E(t)}\\
& \leq & - \alpha \gamma (T(t)-T^\ast)^2 + 
2 \gamma\| A \|_\infty \left| T(t) - T^\ast \right| \sqrt{E(t)},\\
\end{eqnarray*}
where Assumption~\ref{ass:global_assumption} has been used again. Gathering the
estimates,
\begin{eqnarray*}
\frac{d\mathcal{E}(t)}{dt} & \leq & 
 -\rho k_{\rm B} T(t) \, E(t) + \frac{4 \gamma \, \| V \|_\infty}{k_\mathrm{B} T(t)^2}
\sqrt{E(t)} \left ( a \left| T(t) - T^\ast \right| 
+ 2 \| A \|_\infty \sqrt{E(t)} \right ) \\
& & - \alpha \gamma (T(t)-T^\ast)^2 + 
2 \gamma \| A \|_\infty \left| T(t) - T^\ast \right| \sqrt{E(t)} \\
& \leq & -\left(\rho k_{\rm B} T(t) -\frac{8 \gamma \, \| V \|_\infty
  \| A \|_\infty }{k_\mathrm{B} T(t)^2} \right ) E(t) 
- \alpha \gamma (T(t)-T^\ast)^2 \\
& & 
+ \gamma \left ( 2 \| A \|_\infty + \frac{4 a\, \| V
  \|_\infty}{k_\mathrm{B} T(t)^2} \right )
\sqrt{E(t)} \left| T(t) - T^\ast \right| \\
& \leq & -\left[ \rho k_{\rm B} T(t) -\frac{8 \gamma \, \| V \|_\infty \| A \|_\infty }
{k_\mathrm{B} T(t)^2} - \frac{\gamma}{\eta} \left ( \| A \|_\infty + \frac{2 a\, \| V
  \|_\infty}{k_\mathrm{B} T(t)^2} \right ) \right ] E(t) \\
& & - \gamma \left [ \alpha - \eta \left ( \| A \|_\infty + \frac{2 a\, \| V
  \|_\infty}{k_\mathrm{B} T(t)^2} \right ) \right ] (T(t)-T^\ast)^2, \\
\end{eqnarray*}
where the inequality $\dps ab \leq \frac{1}{2\eta} a^2 +
\frac{\eta}{2} b^2$ (for any $\eta > 0$) has been used in
the last step. 
This estimate shows that, with $\eta = \sqrt{\gamma}$ for instance, and choosing
$\gamma$ small enough such that
\[
\rho k_{\rm B} \min(T^A_\mathrm{min},T^\mathrm{LSI}_\mathrm{min})
-\frac{8 \gamma \, \| V \|_\infty \| A \|_\infty }
{k_\mathrm{B} \min(T^A_\mathrm{min},T^\mathrm{LSI}_\mathrm{min})^2} 
- \sqrt{\gamma} \left ( \| A \|_\infty 
+ \frac{2 a\, \| V \|_\infty}{k_\mathrm{B} 
\min(T^A_\mathrm{min},T^\mathrm{LSI}_\mathrm{min})^2} \right ) > 0,
\]
\[
\alpha - \sqrt{\gamma} \left ( \| A \|_\infty + \frac{2 a\, \| V
  \|_\infty}{k_\mathrm{B}
  \min(T^A_\mathrm{min},T^\mathrm{LSI}_\mathrm{min})^2} \right ) > 0,
\]
there exists $\kappa > 0$ such that 
\[
\frac{d\mathcal{E}(t)}{dt} \leq - \kappa \mathcal{E}(t).
\]
Therefore, the entropy decreases, and the initial bounds on the
temperature are satisfied at all times.
Gronwall's lemma finally
shows that $\mathcal{E}(t) \to 0$ exponentially fast.

\begin{remark}[Nonlinear feedback]
\label{remark:NLfeedback}
Convergence results for a nonlinear temperature feedback 
of the form~\eqref{eq:temperatureNLfeedback}
for a nonlinear function $f$ such that $f(x)
= 0$ if and only if $x = 0$,
can be proved by
extending the proofs of this section to the case when 
\begin{equation}
  \label{eq:condition_NL_f}
  \forall R > 0, \quad \exists c_1, c_2 > 0 \ \ \textrm{such that} \ \ \forall
  x \in [-R,R], \quad c_1
  \leq \frac{f(x)}{x} \leq c_2.
\end{equation}
The constant $R$ is related to the maximal value of $\dps
\int_\mathcal{M} A \psi_t$, which is dictated by the initial entropy $\mathcal{E}(0)$:
if $\mathcal{E}(t) \leq \mathcal{E}(0)$, then the temperature $T(t)$
belongs to a bounded set $I_{\mathcal{E}(0)}$, and so, for usual
entropy functions,
\[
\left | \int_{\mathcal{M}} A(q) \psi_t(q) \, dq \right | \leq 
|\langle A \rangle_{T(t)}| + \int_{\mathcal{M}} A |f - 1| \mu_{T(t)} \, dq
\leq \sup_{ T \in I_{\mathcal{E}(0)} } |\langle A \rangle_T|
+ c \sqrt{\mathcal{E}(t)} \leq R,
\]
where $R$ depends only on $\mathcal{E}(0)$, and $c$ depends on the
choice of the entropy function ($c = 2$ for relative entropies). This shows why the bounds
\eqref{eq:condition_NL_f} on some compact interval only are sufficient. For the choice $f(x) = x
+ x^3$ for instance, $c_1 = 1$ and $c_2 = 1+R^2$. Such choices may be
helpful in numerical simulations to accelerate the convergence toward
the temperature of interest at the early stages of the process.
\end{remark}

\subsection{Proof of Theorem~\ref{thm:Poincare_cv}}

The proof follows the same lines as the proof of
Theorem~\ref{thm:LSIcv}, and we present only the modifications
required in the case considered here.
The first change in the proof is in the bound of the derivative
of the spatial entropy $E(t)$.
It holds
\begin{eqnarray*}
\frac{dE(t)}{dt} & = & 
-k_{\rm B} T(t)\int_{\mathcal{M}} \left | \nabla f \right |^2 \, \mu_{T(t)}
+ \frac{T'(t)}{k_\mathrm{B} T(t)^2}
\int_{\mathcal{M}} \left [ h(f) - (f-1) \, f \right ]\, \left(
V - \left \langle V \right \rangle_{T(t)} \right ) \, \mu_{T(t)}\\
& = & -\rho k_{\rm B} T(t) \, E(t)
- \frac{T'(t)}{k_\mathrm{B} T(t)^2}
\int_{\mathcal{M}} \left [ h(f) + f-1 \right ]\, \left(
V - \left \langle V \right \rangle_{T(t)} \right ) \, \mu_{T(t)}\\
& \leq & -\left ( \rho k_{\rm B} T(t) - \frac{2 |T'(t)| \, \| V
  \|_\infty}{k_\mathrm{B} T(t)^2} \right ) E(t) + 
\frac{2 |T'(t)| \| V \|_\infty}{k_\mathrm{B} T(t)^2}
\int_{\mathcal{M}} |f-1| \, \mu_{T(t)}\\
& \leq & -\left ( \rho k_{\rm B} T(t) - \frac{2 |T'(t)| \, \| V
  \|_\infty}{k_\mathrm{B} T(t)^2} \right ) E(t) + 
\frac{2 \sqrt{2} |T'(t)| \| V \|_\infty}{k_\mathrm{B} T(t)^2}
\sqrt{E(t)} \\
\end{eqnarray*}
where we have used the Cauchy-Schwarz inequality
\[
\int_{\mathcal{M}} |f-1| \, \mu_{T(t)} \leq 
\sqrt{\int_{\mathcal{M}} |f-1|^2 \, \mu_{T(t)}} \sqrt{\int_{\mathcal{M}} \mu_{T(t)}}
= \sqrt{2 E(t)}.
\]
A Cauchy-Schwarz inequality is also used to establish a slighty different bound on
the temperature derivative:
\[
|T'(t)| \leq \gamma \left ( a \left| T(t) - T^\ast \right| 
+ \| A \|_\infty \sqrt{2 E(t)} \right ),
\]
and
\[
\frac{d}{dt} \left [ \frac12 (T(t)-T^\ast)^2 \right ] \leq 
 - \alpha \gamma (T(t)-T^\ast)^2 + 
\gamma  \| A \|_\infty \left| T(t) - T^\ast \right| \sqrt{2E(t)}.
\]
The spatial entropy derivative can
then be bounded as
\[
\frac{dE(t)}{dt} \leq  -\left [ \rho k_{\rm B} T(t) - \gamma \frac{4
    \| A \|_\infty\, \| V
  \|_\infty}{k_\mathrm{B} T(t)^2} \left( 1 + \sqrt{\frac{E(t)}{2}}
  \right)  \right ] E(t) 
+ \frac{2 a \gamma \| V \|_\infty}{k_\mathrm{B} T(t)^2}
|T(t) - T^\ast| \left ( E(t) + \sqrt{2E(t)} \right ).
\]
For the total entropy,
\begin{eqnarray*}
\frac{d\mathcal{E}(t)}{dt} & \leq & 
 -\left [ \rho k_{\rm B} T(t) - \gamma \frac{4  \| A \|_\infty\, \| V
  \|_\infty}{k_\mathrm{B} T(t)^2} \left( 1 + \sqrt{\frac{E(t)}{2}} \right) \right ] E(t) 
- \alpha \gamma (T(t)-T^\ast)^2 \\
& & + \sqrt{2} \gamma \left [ \frac{2 a \| V \|_\infty}
  {k_\mathrm{B} T(t)^2} \left( 1 + \sqrt{\frac{E(t)}{2}} \right)
+ \| A \|_\infty \right ] |T(t) - T^\ast| \sqrt{E(t)}.
\end{eqnarray*}
This estimate is analogous to the one obtained in the relative entropy
case, except for some additional factors $\sqrt{E(t)}$, which are
bounded by $\sqrt{\mathcal{E}(t)}$, hence by 
$\sqrt{\mathcal{E}(0)}$ when the total entropy decreases. However, it is
easily seen that, taking initial conditions $\mathcal{E}(0) \leq
\mathcal{E}^\ast $and for $\gamma$
small enough, the function
$\mathcal{E}$ is decreasing and bounded by $\mathcal{E}(0)$. There
exists then $\kappa > 0$ such that 
$\mathcal{E}'(t) \leq - \kappa \mathcal{E}(t)$,
which again implies the longtime convergence.


\subsection*{Acknowledgements}
We thank Eric Canc\`es, Tony Leli\`evre and Fr\'ed\'eric Legoll for
helpful discussions, and the anonymous referee for his stimulating comments.
Part of this work was done while G. Stoltz was participating to the
program ``Computational Mathematics'' at the Haussdorff Institute for
Mathematics in Bonn, Germany.
Support from the ANR INGEMOL of the
French Ministry of Research is acknowledged.


\end{document}